\newif\ifsingle
\singletrue 

\newif\ifproofs

\ifsingle
\documentclass[12pt,draftclsnofoot, onecolumn]{IEEEtran}		
\else		
\documentclass[10pt,final, twocolumn]{IEEEtran}
\fi

\usepackage{times}
\usepackage{amsmath,dsfont}
\usepackage{amssymb,amsthm}
\usepackage{epsfig,verbatim}
\usepackage{subcaption}
\usepackage{setspace}
\usepackage{color}
\usepackage{cite}
\usepackage{epstopdf}
\usepackage{graphics}
\usepackage{accents}
\usepackage{acronym}
\usepackage[bookmarks,colorlinks]{hyperref}
\usepackage{booktabs}
\usepackage{mathtools}
\usepackage{float}
\usepackage[linesnumbered,ruled,lined]{algorithm2e}
\usepackage{algpseudocode}
\usepackage{enumitem}
\usepackage{bm}

\newcommand{\removelatexerror}{\let\@latex@error\@gobble}
\newcommand{\myVec}[1]{{\boldsymbol{#1}}}
\newcommand{\myMat}[1]{{\boldsymbol{#1}}}
\newcommand{\mySet}[1]{\mathcal{#1}}
\newcommand{\E}{\mathds{E}}		 			

\newcommand{\Kiter}{K} 
\newcommand{\Opt}{^{({\rm o})}}

\acrodef{dnn}[DNN]{deep neural network} 
\acrodef{csi}[CSI]{channel state information}
\acrodef{map}[MAP]{maximum a-posteriori probability}
\acrodef{snr}[SNR]{signal-to-noise ratio}
\acrodef{ser}[SER]{symbol error rate}
\acrodef{bs}[BS]{base station} 
\acrodef{ml}[ML]{machine learning} 
\acrodef{iot}[IOT]{Interent of Things}
\acrodef{mimo}[MIMO]{multiple-input multiple-output}
\acrodef{mse}[MSE]{mean-squared error}
\acrodef{pdf}[PDF]{probability density function}
\acrodef{rv}[RV]{random variable}
\acrodef{em}[EM]{expectation maximization}
\acrodef{hmm}[HMM]{hidden Markov model}
\acrodef{pdf}[PDF]{probability density function}
\acrodef{isi}[ISI]{intersymbol interference}  
\acrodef{awgn}[AWGN]{additive white Gaussian noise}
\acrodef{iid}[i.i.d.]{independent and identically distributed}
\acrodef{dl}[DL]{deep learning}
\acrodef{pga}[PGA]{projected gradient ascent}
\acrodef{dma}[DMA]{dynamic metasurface antenna}
\acrodef{pcmp}[PCMP]{projected conceptual mirror prox}
\acrodef{cmp}[CMP]{conceptual mirror prox}
\acrodef{ga}[GA]{gradient ascent}
\acrodef{gd}[GD]{gradient descent}
\acrodef{quadriga}[QuaDRiGa]{Quasi Deterministic Radio channel Generator}

\definecolor{blue}{rgb}{0,0,1}

\DeclareMathOperator*{\argmax}{argmax}

\ifsingle

\newcommand{\figSpace}{\vspace{-0.2cm}}

\else

\newcommand{\figSpace}{\vspace{-0.6cm}}
\fi 

\IEEEoverridecommandlockouts

\title{Learn to Rapidly and Robustly Optimize Hybrid Precoding 
}
\author{
	\IEEEauthorblockN{Ortal Lavi and Nir Shlezinger\\
	} 
	\thanks{
	Parts of this work were presented at the 2022 IEEE Workshop on Signal Processing Advances in Wireless Communications (SPAWC) as the paper \cite{agiv2022learn}.
	This work was supported in part by the Israeli Innovation Authority through the 5G-WIN consortium.
		The authors are with the School of ECE, Ben-Gurion University of the Negev (e-mail: agivo@post.bgu.ac.il; nirshl@bgu.ac.il). 		
	}

	\vspace{-1.0cm}
	
}
\vspace{-0.75cm}

\begin{document}

\maketitle
\pagestyle{plain}
\thispagestyle{plain}
\begin{abstract}
Hybrid precoding plays a key role in realizing massive \ac{mimo} transmitters with controllable cost. \ac{mimo} precoders are required to frequently adapt based on the variations in the channel conditions. In hybrid \ac{mimo}, where precoding is comprised of digital and analog beamforming, such an adaptation  involves lengthy optimization and  depends on accurate \ac{csi}. This affects the spectral efficiency  when the channel varies rapidly and when operating with noisy \ac{csi}. In this work we employ deep learning techniques to learn how to {\em rapidly} and {\em robustly} optimize hybrid precoders, while being fully interpretable. 
We leverage data to learn iteration-dependent hyperparameter settings of projected gradient sum-rate optimization with a predefined number of iterations. The algorithm maps channel realizations into hybrid precoding settings while preserving the interpretable flow of the optimizer and improving its convergence speed. To cope with noisy \ac{csi}, we learn to optimize the minimal achievable sum-rate among all tolerable errors, proposing a robust hybrid precoding based on the projected conceptual mirror prox minimax optimizer. 
Numerical results  demonstrate that our approach allows using over ten times less iterations compared to that required by conventional optimization with shared hyperparameters, while achieving similar and even improved sum-rate performance.
\end{abstract}

\acresetall
\section{Introduction}
Wireless communication networks are subject to constantly growing requirements in terms of connectivity, throughput, and reliability. One of the emerging technologies which is expected to play a key role in meeting these demands is based on equipping wireless \acp{bs} with large-scale antenna arrays, resulting in massive \ac{mimo} networks \cite{samsung202065}. 
While the theoretical gains of massive \ac{mimo} are well-established \cite{marzetta2010noncooperative,bjornson2017massive,shlezinger2018spectral}, implementing such large scale arrays in a power and cost efficient manner is associated with several core challenges. Among these challenges is the conventional need to feed each antenna element with a dedicated RF chain, which tend to be costly and consume notable power~\cite{gao2018low}. 

A leading approach to tackle the cost and power challenges of massive \ac{mimo} is to utilize hybrid analog/digital \ac{mimo} transceivers \cite{molisch2017hybrid,ahmed2018survey}.  
Hybrid \ac{mimo} transceivers carry out part of the processing of the transmitted and received signals in the analog domain, enabling operation with  less RF chains than antennas \cite{mendez2016hybrid,ioushua2019family,gong2019rf}. As a result, hybrid \ac{mimo} transmitters implement precoding partially in digital and partially in analog. Analog processing is dictated by the circuitry, often implemented using vector modulators~\cite{zirtiloglu2022power} or phase shifters~\cite{mendez2016hybrid}. Consequently, analog precoding is typically more constrained compared with  digital processing, where, e.g., one can typically apply different precoders in each frequency~\cite{park2017dynamic}. 

The constrained form of hybrid \ac{mimo} makes the setting of the precoding pattern for a given channel realization notably more challenging compared with costly fully-digital architectures. 
Various methods have been proposed for designing  hybrid precoding systems, optimizing their analog and digital processing to meet the  communication demands \cite{albreem2021overview}. The common approach formulates the objective of the  precoders, e.g., sum-rate maximization or minimizing the distance from the fully-digital precoder~\cite{yu2016alternating,sohrabi2016hybrid}, as an optimization problem. The resulting optimization is then tackled using iterative solvers which vary based on the objective and the specific constraints induced by the analog circuitry and the antenna architecture. Iterative algorithms were proposed for tunning hybrid \ac{mimo} systems with controllable gains \cite{gong2019rf}, phase-shifting structure \cite{ioushua2019family,park2017dynamic},  partial-connectivity \cite{mendez2016hybrid,yu2016alternating}, discretized vector modulators \cite{zirtiloglu2022power}, and  Lorentzian-constrained metasurface antennas \cite{shlezinger2021dynamic,zhang2022beam}. While iterative optimizers are interpretable, being derived as the solution to the formulated problem, they are often  slow in terms of convergence. This can be a major limitation as this setting is based on the instantaneous \ac{csi}, and thus must be done in real-time to cope with the frequent variations of wireless channels, and its performance depends on the accuracy of the \ac{csi}. 

An alternative emerging approach to tuning hybrid precoders is based on \acl{dl}. This approach builds upon the ability of \acp{dnn} to learn complex mapping while inferring at controllable speed dictated by the number of layers, which is often much faster compared with conventional iterative optimizers \cite{zappone2019wireless}. The usage of \acl{dl} to tackle optimization problems is referred to as {\em learn-to-optimize} \cite{chen2021learning}. \ac{dnn}-aided optimization of hybrid precoders was considered in  \cite{hu2021two, huang2019deep, huang2019fast,elbir2019hybrid,elbir2021terahertz,peken2020deep, hojatian2021unsupervised}, which employed multi-layered perceptrons \cite{hu2021two,huang2019deep} and  convolutional neural networks \cite{huang2019fast,elbir2019hybrid,elbir2021terahertz,peken2020deep, hojatian2021unsupervised} as the optimizer, while  \cite{hu2021joint,wang2020precodernet} employed deep reinforcement learning techniques. 
While such deep models are able to map channel estimates into precoder structure, they lack of transparency or interpretability in how input data are transformed into the precoder setting. Furthermore, the resulting precoding scheme is geared towards the number of users and the distribution of the channels used for its training. When the channel distribution or the number of users changes, the \ac{dnn} needs to be trained anew, which is a lengthy procedure. Moreover, highly-parameterized \acp{dnn} may be too computationally complex to  deploy on hardware-limited wireless communication devices.


Interpretable deep models with a small number of parameters can be obtained from iterative optimization algorithms via the deep unfolding methodology \cite{shlezinger2020model}. Deep unfolding leverages data-driven deep learning techniques to improve an iterative optimizer, rather than replace its operation with \acp{dnn}~\cite{shlezinger2022model,balatsoukas2019deep}. Deep unfolded optimizers were utilized for configuring one-bit hybrid precoders in \cite{balatsoukas2019neural}; for fully-digital (non-hybrid) narrowband precoders in  \cite{pellaco2021matrix}; for narrowband phase shifter based hybrid precoders in \cite{luo2022mdl,shi2022deep}; and for  narrowband hybrid predcoders with limited feedback in \cite{balevi2021unfolded}. The latter unrolled an optimization algorithm using generative networks for handling the limited feedback and overparametrized ResNets for optimizing the precoders, resulting in a highly-parameterized \ac{dnn}, which may not be suitable for application on hardware limited \ac{mimo} devices. The work \cite{chen2022hybrid} used dedicated \acp{dnn} integrated into unfolded optimization of analog precoders in hybrid \ac{mimo} transmitters with full \ac{csi}, at the cost of additional complexity during inference.   This motivates designing interpretable light-weight learn-to-optimize algorithms for rapidly translating \ac{csi} into  multi-band hybrid precoders, while being robust to inaccurate \ac{csi}.   

In this work, we propose a learn-to-optimize algorithm for tuning multi-band downlink hybrid \ac{mimo} precoders in a manner that is rapid, robust, and interpretable. We consider different analog architectures, including analog precoders with controllable gains, as well as architectures based on fully-connected phase shifters. Our approach leverages data in the form of past channel realization to accelerate the convergence of conventional  iterative optimization of the achievable sum-rate. 
In order to design an unfolded algorithm for  dealing with the challenging and practical setting where the \ac{csi} may be noisy, we commence by treating the setting where full \ac{csi} is available, for which we formulate an unfolded optimizer which naturally extends to cope with noisy \ac{csi}. 
Our proposed data-aided algorithms fully preserve the interpretable and flexible operation of conventional optimizers from which they originate, while being light-weight and operating with fixed run-time and complexity.

In particular, we adopt the \ac{pga} algorithm as the hybrid precoding designing optimizer for maximizing the achievable sum-rate with full \ac{csi}. Our motivation here stems from the ability of this approach to directly optimize the sum-rate, rather than adopting some simpler surrogate objective as done in, e.g., \cite{yu2016alternating,ioushua2019family}, which can thus be extended to robust optimization of this measure. 
We unfold the \ac{pga} algorithm, by fixing a small number of iterations and train only its iteration-dependent hyperparameters which can be tuned from data without deviating from its overall flow and operation. 

Our selection of \ac{pga} optimization with the sum-rate objective as our basis for hybrid precoding design with full \ac{csi} facilitates extending our approach to noisy \ac{csi} settings: we identify the  \ac{pcmp} algorithm for maximin optimization~\cite{thekumparampil2019efficient} as a suitable optimizer for hybrid precoding with noisy \ac{csi}, whose algorithmic steps involve computations of gradients of projection of the sum-rate objective, as we used by \ac{pga} with full \ac{csi}. Consequently, we propose to cope with noisy \ac{csi} by formulating our design objective as maximizing the minimal achievable sum-rate within a given tolerable \ac{csi} error margin. We then unfold \ac{pcmp}, resulting in a trainable architecture involving similar steps as those used when unfolding \ac{pga},  where again data is leveraged to tune the hyperparameters of the optimizer for each iteration. 
Our experimental results evaluate our  proposed unfolded algorithms compared with the conventional optimizers from which they originate for hybrid precoding in multi-band synthetic Rayleigh channels as well as channels generated via the \ac{quadriga} channel simulator~\cite{jaeckel2014quadriga} in various \acp{snr}. There, we systematically show that the proposed unfolded algorithms achieve similar and even improved sum-rates compared with their iterative counterparts, while operating with over ten times less iterations and computational complexity.


The rest of this work is organized as follows: Section~\ref{sec:System} describes  the system model; the proposed learn-to-optimize algorithm for tuning hybrid precoders with full \ac{csi} is derived in Section~\ref{sec:Hybrid}, while Section~\ref{sec:Hybrid_noise} derives the learned optimizer for noisy \ac{csi}. Our proposed algorithms are numerically evaluated in Section~\ref{sec:Sims}, and Section~\ref{sec:Conclusions} provides concluding remarks.

 Throughout the paper, we use lowercase boldface letters for vectors, e.g., ${\myVec{x}}$; $[{\myVec{x}}]_i$ denotes
the $i$th element of ${\myVec{x}}$. Uppercase boldface letters are used for matrices, e.g., $\myMat{X}$, with $[\myMat{X}]_{i,j}$ being its $(i,j)$th entry and $\myMat{I}_n$ denoting the $n\times n$ identity matrix.
Calligraphic letters, such as $\mySet{X}$, are used for sets, 
and $\mathbb{C}$ is the set of complex numbers.
The transpose, Hermitian transpose, Frobenius norm, and stochastic expectation are denoted by $(\cdot)^T$, $(\cdot)^H$,  $\|\cdot\|_F$, and  $\E\{\cdot\}$, respectively. 
 
\section{System Model}
\label{sec:System}
In this section we present the system model. We start by reviewing the  model for wideband downlink \ac{mimo} communications with hybrid beamforming  in Subsection~\ref{subsec:MIMO}, after which we discuss the considered analog precoder architectures in Subsection~\ref{subsec:Analog}. Then, we formulate the considered problem of rapid and robust hybrid precoder setting in Subsection~\ref{subsec:Problem}.

\subsection{Downlink MIMO Communications with Hybrid Beamforming}
\label{subsec:MIMO}
\subsubsection{Channel Model}
\begin{figure}
    \centering
    \includegraphics[width=\linewidth]{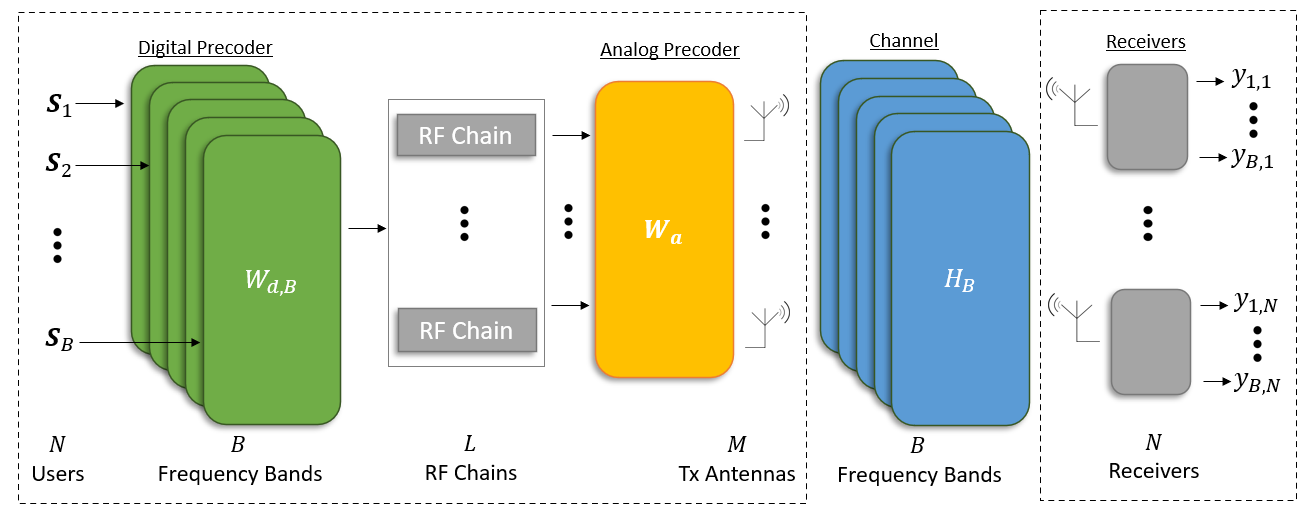}
   \vspace{-0.2cm}
    \caption{Architecture of \ac{mimo} system using hybrid precoding}
    \label{fig:Model1}
\end{figure}

We consider a single-cell downlink hybrid \ac{mimo} system with $N$ single-antenna users. The \ac{bs} is equipped with $M$ transmitting antennas, and utilizes $B$ frequency bands for communications, where the spectrum is shared among all users in a non-orthogonal fashion. The overall system is illustrated in Fig.~\ref{fig:Model1}. 

The \ac{bs} employs multi-carrier signaling, and we use $\textbf{\emph{s}}_b\in\mathbb{C}^N$ to denote the multiuser transmitted signal vector that is being transmitted in the $b$th frequency bin, $b\in \{1,2,\ldots, B\}\triangleq \mySet{B}$. The transmitted symbols are i.i.d. and of equal power, such that  $\mathbb{E} [\textbf{\emph{s}}_b\textbf{\emph{s}}_b^H] = \frac{1}{N}\textbf{I}_N $ for each $b \in \mySet{B}$.
At each frequency bin $b\in \mySet{B}$, for a channel input $\textbf{\emph{x}}_b\in\mathbb{C}^M$, the channel output is given by
	\begin{equation}
	\label{eqn:io relation}
	\textbf{\emph{y}}_b = \textbf{H}_b \cdot \textbf{\emph{x}}_b + \textbf{\emph{n}}_b \in\mathbb{C}^N,
	\end{equation}
where $\textbf{H}_b\in\mathbb{C}^{N\times M}$ is the $b$th frequency band sub-channel, and $\textbf{\emph{n}}_b\in\mathbb{C}^N$ is \ac{awgn} with i.i.d. entries of variance $\sigma^2$.

\subsubsection{Hybrid Beamforming}
The \ac{bs} has $L < M$ RF chains, and thus employs hybrid  precoding. Here, the multiuser transmitted signal vectors $\{\textbf{\emph{s}}_b\}_{b\in \mySet{B}}$ are precoded in two stages. First, a digital precoder $\textbf{W}_{d,b}\in\mathbb{C}^{L\times{N}}$ is applied to $\textbf{\emph{s}}_b$ in each frequency bin $b\in\mySet{B}$, i.e., $\textbf{W}_{d,b}$ is the digital precoding matrix of the $b$th bin. Next, the digital symbols  pass through  $L$ RF chains, and are combined into the channel input $\textbf{\emph{x}}_b $ using an analog precoder. Unlike digital processing,  analog precoding is carried out using dedicated hardware  assumed to be static in frequency. Hence,  analog precoding is modeled using the matrix
$\textbf{W}_a\in \mySet{A} \subseteq \mathbb{C}^{M\times{L}}$, where $\mySet{A}$ represents the set of feasible analog precoder settings, discussed in the sequel. 

The output of the transmitter in the $b$th bin is given by
 \begin{equation}
 \label{eqn:ChInput}
   \textbf{\emph{x}}_b = \textbf{W}_a\textbf{W}_{d,b} \cdot \textbf{\emph{s}}_b.
\end{equation}
We require the precoders to satisfy 
\begin{equation}
\label{eqn:Power}
 \frac{1}{B} \sum_{j=1}^{B}\| \textbf{W}_a\textbf{W}_{d,j} \|^2_F \leq \emph{N},   
\end{equation}
 as the transmitter's total power constraint.
Substituting \eqref{eqn:ChInput} 
into \eqref{eqn:io relation}, the channel output at the $N$ users at frequency bin $b$ can be written as 
	\begin{equation}
	\label{eqn:b received signal}
	\textbf{\emph{y}}_b = \textbf{H}_b\textbf{W}_a\textbf{W}_{d,b} \cdot \textbf{\emph{s}}_b + \textbf{\emph{n}}_b \in\mathbb{C}^N.
	\end{equation}
 
\subsection{Analog Precoders}
\label{subsec:Analog}
The feasible mappings that can be realized by the analog precoder $\textbf{W}_a$, encapsulated in the set $\mySet{A}$, depend on the hardware architecture. We consider two different architectures for the analog precoder: unconstrained precodersZ and fully-connected phase shifter networks.


\subsubsection{Unconstrained Precoder}
Here, the analog precoding matrix, $\textbf{W}_a$, has no hardware constraints, and can realize any complex matrix (while satisfying the overall power constraint \eqref{eqn:Power}). For unconstrained precoders, we use $\mySet{A} = \mathbb{C}^{M\times{L}}$.
Such analog processing with configurable attenuation and phase shifting can be implemented using, e.g., vector modulators, as in \cite{gong2019rf, zirtiloglu2022power}. While this formulation does not assume any constraints, other than the power constraint, it is emphasized that unconstrained solutions can be useful for constrained analog hardware designs, which often involve a preliminary step of obtaining an unconstrained combiner followed by its projection to the constrained set, as done in, e.g.,~\cite{ioushua2019family}.

\subsubsection{Phase Shifter Networks}
\label{subsubsec: phase-shifters-networks}
A common implementation of analog precoders utilizes phase shifters~\cite{ioushua2019family,mendez2016hybrid}. In fully-connected phase shifter networks, every antenna element is connected to each RF chain via a dedicated controllable phase shifter. Such precoders are modelled as matrices whose entries have a unit magnitude, namely,
\begin{equation}
	\label{eqn:ph_shift}
    \mySet{A} = \Big\{\textbf{A}\in \mathbb{C}^{M\times{L}}\Big| \big|[\textbf{A}]_{m,l}\big| = 1, \quad \forall (m,l) \Big\}.
\end{equation}

\subsection{Problem Formulation}
\label{subsec:Problem}
We aim at designing the hybrid precoding operation given a channel realization to maximize the achievable sum-rate of multiuser downlink \ac{mimo} communications. Defining the sum-rate as the precoding objective is a common approach \cite{zhang2022beam, gao2016energy, park2017dynamic}, being a communication measure of the overall combined effect of the channel and the hybrid precoder. We particularly focus on two scenarios -- sum-rate maximization (given accurate \ac{csi}) and robust sum-rate maximization (given noisy \ac{csi}). 

\subsubsection{Sum-Rate Maximization}
Since $[\textbf{\emph{s}}_b\textbf{\emph{s}}_b^H] = \frac{1}{N}\textbf{I}_N $, it holds that the following sum-rate is achievable~\cite{shen2007sum}
	\begin{align}
	\emph{R}\left(\textbf{W}_a , \{\textbf{W}_{d,b}\}_{b\in \mySet{B}}, \{\textbf{H}_b\}_{b\in \mySet{B}}\right) 
	 = \frac{1}{B} \sum_{b=1}^{B} \log \left|\textbf{I}_N \!+\! \frac{1}{N \sigma^2} \textbf{H}_b\textbf{W}_a\textbf{W}_{d,b} \textbf{W}_{d,b}^H\textbf{W}_a^H\textbf{H}_b^H\right|.
	\label{eqn:achievable rate}
	\end{align}
Consequently, for a given channel realization $\{\textbf{H}_b\}_{b\in \mySet{B}}$, we aim to find $\textbf{W}_a\Opt$ and $\{\textbf{W}_{d,b}\Opt\}_{b\in \mySet{B}}$, that are the solution to the following optimization problem
	\begin{align}
	\label{eqn:main problem}
	\left(\textbf{W}_a\Opt,\{\textbf{W}_{d,b}\Opt\}_{b\in \mySet{B}}\right) &=\argmax\limits_{\textbf{W}_a , \{\textbf{W}_{d,b}\}_{b\in \mySet{B}}} R\left(\textbf{W}_a , \{\textbf{W}_{d,b}\}, \{\textbf{H}_b\}\right) 
	\\ &{\rm s.t.}~~ 
	\begin{cases} 
      \frac{1}{B} \sum_{b=1}^{B}\| \textbf{W}_a\textbf{W}_{d,b} \|^2_F \leq \emph{N} \\
      \textbf{W}_a \in \mySet{A}
   \end{cases}\notag.
	\end{align}
In \eqref{eqn:main problem}, $\mySet{A}$ is the set of matrices corresponding to the analog precoding models described above, i.e., we seek to find a suitable precoders under the different constraints detailed in Subsection~\ref{subsec:Analog}.

Since the optimization problem \eqref{eqn:main problem} must be tackled each time the channel realization changes, our proposed solution should not only tackle \eqref{eqn:main problem}, but also to do it rapidly, i.e., with a predefined amount of computations. To achieve this aim, we assume access to a set $\mathcal{D}$ of previously encountered or simulated channel realizations $\{{\textbf{H}}^r\}_{r=1}^{|\mathcal{D}|}$, which can be exploited to facilitate optimization within a fixed number of operations (dictated by, e.g., the coherence duration of the channel).

\subsubsection{Robust Sum-Rate Maximization}
The optimization problem in \eqref{eqn:main problem} relies on knowledge of the channel in each frequency bin, i.e., on full \ac{csi}. 
Using \eqref{eqn:achievable rate} as our objective is expected to affect the performance of the hybrid precoders when the channel matrices provided as inputs are (possibly inaccurate) estimates of the true channel. Consequently, we seek to optimize the hybrid precoders in a manner that is robust to a predefined level of inaccuracies in the \ac{csi}. 

To formulate this, we consider a noisy sub-channel estimation denoted $\left\{\textbf{H}_b + \textbf{E}_b\right\}_{b\in\mySet{B}}$, where $\textbf{H}_b$ is the true channel realization, and $\textbf{E}_b$ is the estimation error. We wish to design the hybrid precoders to be robust to estimation errors within a predefined level $\varepsilon$, i.e., when $\| \textbf{E}_b \|_F < \varepsilon$ for each $b\in\mathcal{B}$. Consequently, we convert \eqref{eqn:main problem} into a maximin problem
	\begin{align}
	\label{eqn:max min problem}
	\left(\textbf{W}_a\Opt,\{\textbf{W}_{d,b}\Opt\}_{b\in \mySet{B}}\right) &=\argmax\limits_{\textbf{W}_a , \{\textbf{W}_{d,b}\}_{b\in \mySet{B}}} \left\{ \min_{\| \textbf{E}_b \|_F < \varepsilon} R\left(\textbf{W}_a , \{\textbf{W}_{d,b}\}, \left\{\textbf{H}_b + \textbf{E}_b\right\}\right) \right\}
	\\ &{\rm s.t.}~~ 
    	\begin{cases} 
          \frac{1}{B} \sum_{b=1}^{B}\| \textbf{W}_a\textbf{W}_{d,b} \|^2_F \leq \emph{N} \\
          \textbf{W}_a \in \mySet{A}
       \end{cases}\notag .
	\end{align}
In \eqref{eqn:max min problem}, we seek to maximize the minimal rate resulting from a tolerable estimation error of the channel, i.e., an estimation error within the predefined level $\varepsilon$. Again, we wish to carry out robust optimization rapidly, and can leverage the set of past channel realizations $\mathcal{D}$ to that aim.


\section{Rapid Hybrid Precoder Learned Optimization with Full \ac{csi}}
\label{sec:Hybrid}
We first consider the rapid tuning of the hybrid precoder with full \ac{csi}. 
For each realization of the wireless channel $\{\textbf{H}_b\}_{b\in \mySet{B}}$, the configuration of the hybrid precoder can be formulated as a constrained optimization problem \eqref{eqn:main problem}. Consequently, one can design a hybrid precoder using suitable iterative optimization methods. A candidate method for tackling the optimization problem in \eqref{eqn:main problem} is \ac{pga}. \ac{pga} can directly tackle the multi-carrier sum-rate objective in \eqref{eqn:main problem}, as opposed to alternative iterative optimizers for hybrid beamforming which use a relaxed objective aiming to approach the fully-digital beamformer as in \cite{yu2016alternating, ioushua2019family}. An additional motivation for aiming at directly optimizing the sum-rate rather than an alternative surrogate objective (which may be desired to optimize) stems from the fact that the resulting derivation can be extended to robust optimization, i.e., to cope with noisy \ac{csi}, as detailed in Section~\ref{sec:Hybrid_noise}.
Setting the hybrid precoder setup is a channel-dependent task. Consequently, one would have to carry out the optimization procedure each time the channel changes, i.e., on each coherence duration, while principled iterative optimizers such as \ac{pga} tend to be slow and lengthy. 
To cope with this challenge, in this section we present the method of learn-to-optimize hybrid precoding with full \ac{csi}, which is specifically designed to rapidly configure hybrid precoders. Our scheme is based on the application of \ac{pga} to \eqref{eqn:main problem},
as detailed in Subsection~\ref{subsec:PGA}, whose convergence speed we optimize without compromising on interpretability and suitability using deep unfolding \cite{shlezinger2022model}, as we derive in Subsection~\ref{subsec:SymbolDetDeep} and discuss in Subsection~\ref{subsec:Discussion}.


\subsection{Projected Gradient Ascent}
\label{subsec:PGA}
Problem \eqref{eqn:main problem} represents constrained maximization with $B+1$ optimization matrix variables  $\textbf{W}_a, \{\textbf{W}_{d,b}\}_{b\in \mySet{B}}$. Such problems can be tackled using the \ac{pga} algorithm combined with alternating optimization. In this iterative method, each iteration first optimizes $\textbf{W}_a$ while keeping $\{\textbf{W}_{d,b}\}_{b\in \mySet{B}}$ fixed, then repeats this process for every $\textbf{W}_{d,b}$. The optimized matrices are projected to guarantee the constraints are not violated.

To formulate this operation mathematically, we define $\tilde{\textbf{H}}_b \triangleq \sqrt{\frac{1}{N \sigma^2}}\textbf{H}_b$. Each iteration of index $k+1$ is comprised of two alternating stages: analog \ac{pga} and digital \ac{pga}.

{\bf Analog \ac{pga}}:  
The update of $\textbf{W}_a$ is carried out according to 
	\begin{equation} 
	\textbf{W}_a^{(k+1)} 
	= \Pi_{\mySet{A}}\left\{\textbf{W}_a^{(k)}
	+ \mu_a^{(k)}\frac{\partial}{\partial\textbf{W}_a}
	\emph{R}(\textbf{W}_a^{(k)}, \{\textbf{W}_{d,b}^{(k)}\}, \{\textbf{H}_b\})\right\}, 
	\label{eqn:Wa_update}
	\end{equation}
where $\mu_a^{(k)}$ is the step size of the gradient step, and $\Pi_{\mySet{A}}\{\cdot\}$ is the projection operator onto the set $\mySet{A}$. The gradient of $\emph{R}$ with respect to $\textbf{W}_a$ is given by (see Appendix \ref{sec:dev Wa grad}) 
	\begin{align}
	\frac{\partial}{\partial\textbf{W}_a}
	\emph{R}(\textbf{W}_a  , \{\textbf{W}_{d,b}\}, \{\textbf{H}_b\} ) 
	= \frac{1}{B} \sum_{b=1}^{B}\tilde{\textbf{H}}_b^T\textbf{G}_b(\textbf{W}_a ,\textbf{W}_{d,b} , \textbf{H}_b)^{-T} \tilde{\textbf{H}}_b^*
	\textbf{W}_a^{*} \textbf{W}_{d,b}^{*} \textbf{W}_{d,b}^T,
	\label{eqn:gr_Wa_k} 
	\end{align}
where  $\textbf{G}_b(\textbf{W}_a,\textbf{W}_{d,b}, \textbf{H}_b) \triangleq  (\textbf{I}_N +  \tilde{\textbf{H}}_b\textbf{W}_a\textbf{W}_{d,b} \textbf{W}_{d,b}^{H}\textbf{W}_a^H \tilde{\textbf{H}}_b^H )$. 

The projection operator $\Pi_{\mySet{A}}$ in \eqref{eqn:Wa_update} depends on the feasible analog mappings. For unconstrained architectures, clearly $\Pi_{\mySet{A}}(\textbf{A})=\textbf{A}$. For fully connected phase shifters based hardware, the projection is given by
	\begin{align}
	\label{eqn:pr_Wa_partially} 
	&\Pi_{\mySet{A}}\{\textbf{A}\} = \tilde{\textbf{A}}, 
	\qquad [\tilde{\textbf{A}}]_{m,l} = \frac{[\textbf{A}]_{m,l}}{|[\textbf{A}]_{m,l}|},  \forall (m,l).  
	\end{align}
{\bf Digital \ac{pga}}:  
The digital precoder is updated after the analog precoder, where the update is carried out for all frequencies in parallel. Namely, the gradient step for each $ b\in \mySet{B}$ is computed as
	\begin{align}
	\hat{\textbf{W}}_{d,b}^{(k+1)} 
	= \textbf{W}_{d,b}^{(k)}
+ \mu_{d,b}^{(k)}\frac{\partial}{\partial\textbf{W}_{d,b}}
	\emph{R}\Big(&\textbf{W}_a^{(k+1)} , \{\textbf{W}_{d,b}^{(k)}\}, \{\textbf{H}_b\} \Big), 
	\label{eqn:Wd_k_hat} 
	\end{align}
where $\mu_{d,b}^{(k)}$ is the step size. The gradient of $\emph{R} $ with respect to $\textbf{W}_{d,b}$ for each $b \in \mySet{B}$ is computed as (see  Appendix \ref{sec:dev Wa grad})
	\begin{align}
	\frac{\partial  }{\partial\textbf{W}_{d,b}}	\emph{R}(\textbf{W}_a, \{\textbf{W}_{d,b}\}, \{\textbf{H}_b\}) = \frac{1}{B}\textbf{W}_a^T\tilde{\textbf{H}}_b^T\textbf{G}_b(\textbf{W}_a,\textbf{W}_{d,b}, \textbf{H}_b)^{-T}\tilde{\textbf{H}}_b^* \textbf{W}_a^{*} \textbf{W}_{d,b}^{*}. \label{eqn:gr_Wd_k}
	\end{align}

After the gradient step is taken for all frequency bins, the solution is projected to meet the power constraint via
	\begin{equation}
	\label{eqn:pr_Wd_k} 
	\textbf{W}_{d,b}^{(k+1)} = \sqrt{\frac{NB}{\sum_{b=1}^{B}\| \textbf{W}_a^{(k+1)}\hat{\textbf{W}}_{d,b}^{(k+1)} \|^2_F}} \cdot \hat{\textbf{W}}_{d,b}^{(k+1)}. 
	\end{equation}


The overall procedure is summarized as 	Algorithm~\ref{alg:PGA}. While the initial settings of  $\{\textbf{W}_{d,b}^{(0)}\}_{b\in \mySet{B}}$ are taken to be random, the initial analog combiner $\textbf{W}_a^{(0)}$ is set to be the first $L$ right-singular vectors of $\frac{1}{B} \sum_{b=1}^{B}\tilde{\textbf{H}}_b$ (the frequency bands sub-channels average). This setting corresponds to analog beamforming towards the eigenmodes of the (frequency-average) channel, being the part of the capacity achieving precoding method for frequency flat \ac{mimo} channels  \cite[Ch. 10]{goldsmith2005wireless}. This principled initialization is numerically shown to notably improve the performance of hybrid precoders tuned to directly optimize the rate in \eqref{eqn:main problem}. 

  \begin{algorithm}
    \caption{Projected Gradient Ascent for Hybrid Precoding}
    \label{alg:PGA}
    \SetAlgoLined
    \SetKwInOut{Initialization}{Init}
    \Initialization{Randomize $\{\textbf{W}_{d,b}^{(0)}\}_{b\in \mySet{B}}$ \newline
     $\textbf{W}_a^{(0)} \leftarrow$ first  $L$ right-singular vectors of $\frac{1}{B} \sum_b\tilde{\textbf{H}}_b$ \newline Set step sizes $\{\mu_{d,b}^{(k)}\}_{b\in \mySet{B}},\mu_a^{(k)}$ }
    \SetKwInOut{Input}{Input}
    \Input{Channel matrices $\{\tilde{\textbf{H}}_b\}_{b\in \mySet{B}}$}  
    {
        \For{$k = 0, 1, \ldots$ until convergence}{%
                    
                    Update $\textbf{W}_a^{(k+1)}$ via \eqref{eqn:Wa_update} \label{line:P_Wa}
                    
                    \For{$b = 1, \ldots, B$}{%
                    Calculate $\hat{\textbf{W}}_{d,b}^{(k+1)}$ by \eqref{eqn:Wd_k_hat} 
                    
                    }
                    \For{$b = 1, \ldots, B$}{%
                    Update $\textbf{W}_{d,b}^{(k+1)}$ via \eqref{eqn:pr_Wd_k} \label{line:P_Wd}
                    }
                
                }
        \KwRet{$\{\textbf{W}_{d,b}^{(k)}\}_{b\in \mySet{B}}$ and $\textbf{W}_a^{(k)}$}
  }
\end{algorithm}

The convergence speed of gradient-based optimizers largely depends on the step sizes, i.e.,  $\big\{\big(\big\{\mu_{d,b}^{(k)}\big\}_{b\in \mySet{B}},\mu_a^{(k)} \big)\big\}$ in Algorithm~\ref{alg:PGA}. However, conventional step size optimization methods based on, e.g., line search and backtracking \cite[Ch. 9]{boyd2004convex}, typically involve additional per-iteration processing which increases the overall complexity and run-time. Hence, a common practice is to use pre-defined hand-tuned constant step sizes, which may result in lengthy convergence.

\subsection{Learn-to-Optimize Hybrid Precoding}
\label{subsec:SymbolDetDeep}
Algorithm~\ref{alg:PGA} optimizes hybrid precoders for a given channel realization. However, its convergence speed largely depends on its hyperparameters, i.e., the step sizes, which tend to be difficult to set. 
Here, we propose to leverage automated data-based optimizers used in \acl{dl} to tune iteration-dependent step sizes, i.e., to learn-to-optimize hybrid precoders in a small and predefined number of iterations. 

Our design follows the deep unfolding methodology \cite{shlezinger2020model, shlezinger2022model}, which designs \acp{dnn} as iterative optimizers with a fixed number of iterations. In particular, we use as our optimizer the \ac{pga} method in Algorithm~\ref{alg:PGA} with exactly $\Kiter$ iterations. By doing so, we guarantee an exact pre-known, and typically small, run-time and complexity in real-time. While the accuracy of first-order optimizers such as \ac{pga} is typically invariant of the setting of the hyperparameters when allowed to run until convergence (under mild conditions, e.g., that the step sizes are sufficiently small), their performance is largely affected by these values when the number of iterations is fixed. Consequently, our design treats the hyperparameters of an iterative optimizer with $\Kiter$ iterations as the parameters of a \ac{dnn} with $\Kiter$ layers, and tunes them  via end-to-end training, based on the available data set $\mySet{D}$, thus converting \ac{pga} into a trainable discriminative model~\cite{shlezinger2022discriminative}.   

To formulate this, the step sizes vector of the $k$th iteration is defined as $\bm{\mu}_{k} \triangleq  (\mu_a^{(k)} ,  \ldots, \mu_{d,B}^{(k)})$, and the step sizes matrix defined as $\bm{\mu} \triangleq (\bm{\mu}_{0},  \ldots, \bm{\mu}_{\Kiter-1})^T$ for $\Kiter$ iterations of Algorithm~\ref{alg:PGA}. The entries of this $\Kiter \times (B+1)$ matrix are trainable parameters, that are learned from data. Note that end-to-end training is feasible despite the fact that $\mySet{D}$ does not hold the ground truth precoders. This follows since the performance of hybrid precoders can be evaluated using the differentiable measure in  \eqref{eqn:achievable rate}, that is used to define a loss function with which the hyperparameters $\bm{\mu}$ are trained in an unsupervised manner. 

The loss function, for a given normalized  channel $\tilde{\textbf{H}} = \{\tilde{\textbf{H}}_b\}_{b\in \mySet{B}}$ and step sizes $\bm{\mu}$, is computed as a weighted average of the negative resulting achievable sum-rates of this channel \eqref{eqn:achievable rate} in each $\textbf{PGA}_{\Kiter}(\tilde{\textbf{H}}, \bm{\mu})$ iteration. This implies that the $k$th iteration sum-rate is computed when the precoders are $\Big(\textbf{W}_a^{(k)}, \{\textbf{W}_{d,b}^{(k)}\}_{b\in \mySet{B}}\Big) \triangleq \textbf{PGA}_{k}(\tilde{\textbf{H}}, \bm{\mu}) $, i.e., the precoders obtained via Algorithm~\ref{alg:PGA} after $k < \Kiter$ iterations and step sizes $\bm{\mu}$. Since each iteration is required to provide a setting of the hybrid beamformer which gradually improves along the iterative procedure, we adopt the following loss, inspired by \cite{samuel2019learning}   
\begin{align}
\label{eqn:loss}
     \mathcal{L}(\tilde{\textbf{H}}, \bm{\mu})  =  \frac{1}{K}\sum_{k=1}^{K} \log(1+k) \cdot \biggl(\frac{-1}{B}  \sum_{b=1}^{B}  \log \left|\textbf{I}_N   +  \tilde{\textbf{H}}_b\textbf{W}_a^{(k)}\textbf{W}_{d,b}^{(k)} \big(\tilde{\textbf{H}}_b\textbf{W}_a^{(k)}\textbf{W}_{d,b}^{(k)}\big)^H\right|\biggl).
\end{align}
The data set $\mySet{D}$ includes channel realizations. Since $\tilde{\textbf{H}}_b \triangleq \sqrt{\frac{1}{N \sigma^2}}\textbf{H}_b$ with known $N$ and $\sigma^2$, we henceforth write the entries of the data set $\mySet{D}$ as $\{\tilde{\textbf{H}}^r\}_{r=1}^{|\mathcal{D}|}$. 

\begin{figure}
    \centering
    \includegraphics[width=\columnwidth]{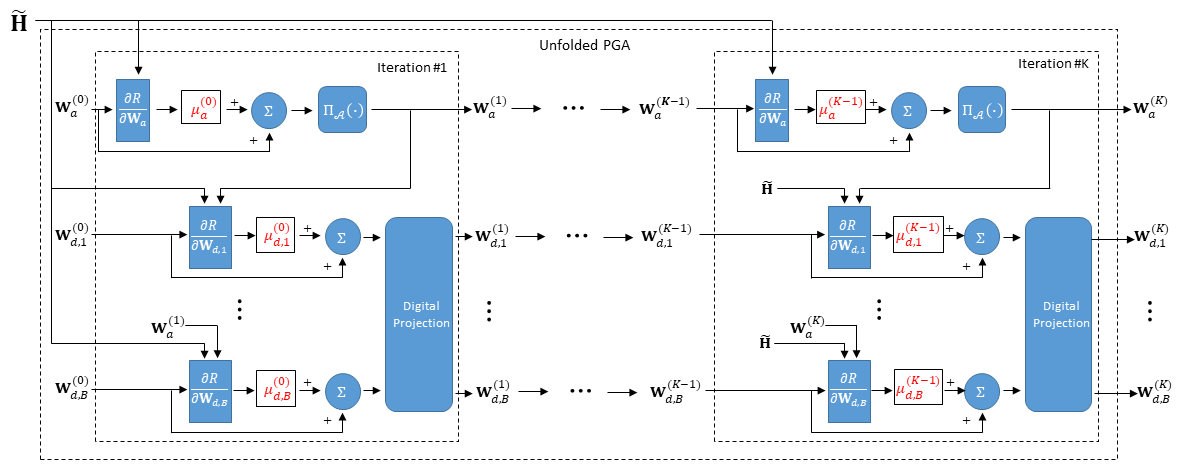}
    \figSpace
    \caption{Unfolded PGA block illustration. The learned parameters are marked in red fonts.}
    \label{fig:unfold pga}
\end{figure}

The learn-to-optimize method seeks to tune the hyperparameters vector $\bm{\mu}$ to best fit the data set $\mathcal{D}$ in the sense of the loss measure \eqref{eqn:loss}. Namely, we aim at setting
\begin{equation}
\label{eqn:L2Oobj}
    \bm{\mu}\Opt = \mathop{\arg \min}\limits_{\bm{\mu}} \frac{1}{|\mathcal{D}|} \sum_{r =1}^{|\mathcal{D}|}\mathcal{L}(\tilde{\textbf{H}}^r, \bm{\mu}).
\end{equation}
We tackle \eqref{eqn:L2Oobj}
using \acl{dl} optimization techniques based on, e.g., mini-batch stochastic gradient descent, to tune $\bm{\mu}$ based on the data set $\mySet{D}$. The resulting procedure is summarized as Algorithm~\ref{alg:L2O}.
We initialize $\bm{\mu}$ before the training process, with fixed step sizes with which \ac{pga} converges. 
After training, which is based on past channel realization and can thus be done offline, the learned $\bm{\mu}$ is used as hyperparameters for rapidly converting a channel realization into a hybrid precoding setting via $\Kiter$ of Algorithm~\ref{alg:PGA}. The resulting unfolded \ac{pga} algorithm is illustrated in Fig.~\ref{fig:unfold pga}.

  \begin{algorithm}
    \caption{Learn-to-Optimize Hybrid Precoding with Full \ac{csi}}
    \label{alg:L2O}
    \SetAlgoLined
    \SetKwInOut{Initialization}{Init}
    \Initialization{Set $\bm{\mu}$ as fixed step sizes. \newline Fix learning rate $\eta$}
    \SetKwInOut{Input}{Input}
    \Input{Training set  $\mathcal{D} = \{\tilde{\textbf{H}}^r\}_{r=1}^{|\mathcal{D}|}$}  
    {
        \For{${\rm epoch} = 0, 1, \ldots, {\rm epoch}_{\max}-1$}{%
                    Randomly divide  $\mathcal{D}$ into $Q$ batches $\{\mathcal{D}_q\}_{q=1}^Q$
                    
                    \For{$q = 1, \ldots, Q$}{
                    
                    Compute precoders via $\textbf{PGA}_{\Kiter}(\mathcal{D}_q , \bm{\mu}) $
                    
                    Compute the average loss of the batch: $\mathcal{L}(\bm{\mu}) = \frac{1}{|\mathcal{D}_q|} \sum_{\tilde{\textbf{H}} \in \mathcal{D}_q}\mathcal{L}(\tilde{\textbf{H}}, \bm{\mu})$
                    
                    Update  $\bm{\mu}\leftarrow \bm{\mu} - \eta\nabla_{\bm{\mu}}\mathcal{L}(\bm{\mu})$ \label{stp:update1}
                    }
                    
                    }
        \KwRet{$\bm{\mu}$}
  }
\end{algorithm}

\subsection{Discussion}
\label{subsec:Discussion}
 The proposed learn-to-optimize method leverages data to improve the performance and convergence speed of iterative \ac{pga}-based optimization. The resulting precoder design  preserves the interpretability and simplicity of classic \ac{pga} optimization, while inferring at a fixed and low delay as done by \acp{dnn} applied for such tasks. We thus benefit from the best of both worlds of model-based optimization and data-driven deep learning.
 
 The fact that the number of iterations is fixed and limited is reflected upon the computational complexity associated with setting a hybrid precoder for a given channel realization. To quantify the complexity, we examine one iteration of Algorithm~\ref{alg:PGA}; its complexity is dominated by the gradient computation (whose complexity order is 
 $ \mySet{O}\bigl(B M^2 ( N+L )\bigl)$) in Step~\ref{line:P_Wa} (which is of the same complexity order as the $B$ gradient computations in Step~\ref{eqn:Wd_k_hat}), and by the projection required in Step~\ref{line:P_Wd} (whose complexity order is $ \mySet{O}\bigl(B N M L\bigl)$). For $\Kiter$ iterations of Algorithm~\ref{alg:PGA} and when signaling over $B$ frequency bands, it follows that the overall complexity of the \ac{pga} algorithm with pre-defined $\Kiter$ iterations, as is done by the proposed learned optimizer, is of the order of 
 \begin{equation}
 \label{eqn:CompPGA}
  \mySet{C}_{\rm PGA} = \mySet{O}\Bigl(\Kiter \cdot B M \bigl(  N  L + M (N+L) \bigl) \Bigl).   
 \end{equation}
 The fact that we set our objective to be the rate $R(\cdot)$ results in its gradient computation yielding a higher complexity per iteration compared with that used when taking the gradients of a surrogate objective, e.g., \cite{yu2016alternating} (where the quadratic dependence is on the number of RF chains rather than the number of antennas), while being of a similar order of that used in \cite{sohrabi2016hybrid}. However, recall that our derivation of Algorithm~\ref{alg:L2O} serves as the first step towards rapid and {\em robust} optimization of the rate, for which these gradient computations are useful, as shown in the sequel. Furthermore, the proposed learn-to-optimize framework facilitates implementing the optimizer with a fixed and small number of iterations, allowing to limit the overall complexity.

 Our approach follows the deep unfolding methodology \cite{shlezinger2020model,balatsoukas2019deep,shlezinger2022model}. As opposed to other forms of deep unfolded networks, which designed \acp{dnn} to imitate the operation of a model-based optimizer while modifying its operation, as  in, e.g., \cite{samuel2019learning}, our design is geared to preserve the operation of model-based iterative optimization. We use automated training capabilities of \acl{dl} tools to tune the hyperparameters of the optimizer. By doing so, we improve upon the conventional usage of fixed hyperparameters, as demonstrated in Section~\ref{sec:Sims}, and avoid the excessive delay of implementing hyperparameter search in each iteration. 
 
 The design objective used in our derivation is the achievable sum-rate \eqref{eqn:achievable rate}, which is approached using dedicated coding over asymptotically large blocks. While one is likely to deviate from \eqref{eqn:achievable rate} in practice, it serves as a fundamental characteristic of the overall channel which encompasses hybrid precoding, transmission, and reception, making it a relevant figure-of-merit for optimizing hybrid \ac{mimo} systems. Furthermore, as the objective in \eqref{eqn:achievable rate} is explicit and differentiable, it enables our method to learn-to-optimize in an unsupervised manner, i.e., one does not need access to ground-truth precoders as in \cite{zappone2019wireless}, and can train using solely channel realizations as data. Yet, using \eqref{eqn:achievable rate} as our objective is expected to affect the performance of the hybrid precoders when the channel matrices provided as inputs are inaccurate estimates of the true channel, in the same manner that such \ac{csi} errors affect the performance of model-based optimizers. We discuss this extension of our method to handling \ac{csi} errors in the following section.

 
 
 
 

\section{Rapid and Robust Hybrid Precoder Learned Optimization with Noisy \ac{csi}}
\label{sec:Hybrid_noise}
In Section~\ref{sec:Hybrid}, we designed the hybrid precoder assuming full \ac{csi}. The resulting Algorithm~\ref{alg:L2O} is compatible for a specific channel, therefore, when the estimation of the channel is noisy, a performance degradation is expected. When dealing with mismatched \ac{csi}, the optimization problem can be formulated as \eqref{eqn:max min problem}, where the {\em minimal rate over all bounded errors} is maximized. In this section we present a method which builds upon our derivation of Algorithm~\ref{alg:L2O} for tackling \eqref{eqn:max min problem} via rapid optimization with a fixed run-time. 
Our approach is based on the \ac{pcmp} algorithm~\cite{thekumparampil2019efficient}, which is an iterative optimizer suitable for maximin objectives as in \eqref{eqn:max min problem}. As detailed in Subsection~\ref{subsec:RobustOpt}, \ac{pcmp} relies on projected gradient steps, and thus its derivation can utilize steps obtained for non-robust optimization via \ac{pga}. Thus, following the approach used in Section~\ref{sec:Hybrid}, we employ deep unfolding, leveraging data to optimize its performance within a fixed number of iterations, as described in Subsection~\ref{subsec:rob l2o}. Then, we provide a discussion in Subsection~\ref{subsec:Discussion2}.

\subsection{Projected Conceptual Mirror Prox Robust Optimization}
\label{subsec:RobustOpt}
The formulation in 
\eqref{eqn:max min problem} represents a constrained maximin optimization problem with $2B+1$ optimization (and auxiliary) matrix variables $\textbf{W}_a, \{\textbf{W}_{d,b}\}_{b\in \mySet{B}}, \{\textbf{E}_b\}_{b\in \mySet{B}}$. Such problems can be tackled by combining the \ac{cmp} algorithm \cite{thekumparampil2019efficient} with alternating optimization and projections, resulting in the \ac{pcmp} method. 



\ac{pcmp} is an iterative method. 
Each iteration is comprised of two stages: \ac{cmp} and projection. We next formulate these stages in the context of \eqref{eqn:max min problem}.

{\bf \ac{cmp}}: The objective in \eqref{eqn:max min problem} is maximized according to the analog and digital precoding matrices, and minimized according to the error matrices. 
\ac{cmp} aims at iteratively refining the optimization variables via gradient steps, which at iteration index $k+1$ take  the form
\begin{align}
    &\left( \textbf{W}_a^{(k+1)}, 
        \{\textbf{W}_{d,b}^{(k+1)}\}_{b\in \mySet{B}}, \{\textbf{E}_b^{(k+1)}\}_{b\in \mySet{B}} \right) = \left( \textbf{W}_a^{(k)}, 
        \{\textbf{W}_{d,b}^{(k)}\}_{b\in \mySet{B}}, \{\textbf{E}_b^{(k)}\}_{b\in \mySet{B}} \right) \notag \\
        &\qquad +\mu^{(k)}\left( 
        \frac{\partial}{\partial\textbf{W}_a},
         \left\{ \frac{\partial}{\partial\textbf{W}_{d,b}}\right\},
         -\left\{ \frac{\partial}{\partial\textbf{E}_{b}}\right\}
        \right) 	\emph{R} \left( \textbf{W}_a^{(k+1)}, 
        \{\textbf{W}_{d,b}^{(k+1)}\}_{b\in \mySet{B}}, \{\textbf{H}_b + \textbf{E}_b^{(k+1)}\}_{b\in \mySet{B}} \right).
        \label{eqn:CMP1}
\end{align}

The computation in \eqref{eqn:CMP1} cannot be directly implemented in general since the gradients are taken with respect to the updated optimization variables (i.e., the iteration index $k+1$ appears on both sides of the update equation). However, it can be approached via additional iterative updates with index $i$ of the form \cite[Eq. (6)]{thekumparampil2019efficient}
\begin{subequations}
\label{eqn:cmp} 
	\begin{align}
            \hat{\textbf{W}}_{a,(i)} &= \textbf{W}_a^{(k)}
        	+ \mu_{a,(i)}^{(k)}\cdot\frac{\partial}{\partial\textbf{W}_a}
        	\emph{R}\left( \hat{\textbf{W}}_{a,(i-1)}, 
            \{\hat{\textbf{W}}_{d,b,(i-1)}\},\{\textbf{H}_b +  \hat{\textbf{E}}_{b,(i-1)}\} \right), \\
            \hat{\textbf{W}}_{d,b,(i)} &= \textbf{W}_{d,b}^{(k)}
        	+ \mu_{d,b(i)}^{(k)}\cdot\frac{\partial}{\partial\textbf{W}_{d,b}}
        		\emph{R}\left( \hat{\textbf{W}}_{a,(i-1)}, 
            \{\hat{\textbf{W}}_{d,b,(i-1)}\},\{\textbf{H}_b +  \hat{\textbf{E}}_{b,(i-1)}\} \right) , \forall b\in \mySet{B}, \\
            \hat{\textbf{E}}_{b,(i)} &= \textbf{E}_b^{(k)}
        	- \mu_{e,b(i)}^{(k)}\cdot\frac{\partial}{\partial\textbf{E}_b}
        		\emph{R}\left( \hat{\textbf{W}}_{a,(i-1)}, 
            \{\hat{\textbf{W}}_{d,b,(i-1)}\},\{\textbf{H}_b + \hat{\textbf{E}}_{b,(i-1)}\} \right) , \forall b\in \mySet{B}. 
	\end{align}
\end{subequations}	
In \eqref{eqn:cmp}, $\mu_{a(i)}^{(k)}, \mu_{d,b(i)}^{(k)}, \mu_{e,b(i)}^{(k)}$ are the step sizes.  The optimization variables in \eqref{eqn:cmp} are initialized to
	\begin{align}
	\label{eqn:init}
        \left( \hat{\textbf{W}}_{a,(0)}, 
        \{\hat{\textbf{W}}_{d,b,(0)}\}_{b\in \mySet{B}}, \{\hat{\textbf{E}}_{b,(0)}\}_{b\in \mySet{B}} \right) = 
        \left( \textbf{W}_a^{(k)}, 
        \{\textbf{W}_{d,b}^{(k)}\}_{b\in \mySet{B}}, \{\textbf{E}_b^{(k)}\}_{b\in \mySet{B}} \right) .
	\end{align}
	
The computation of the gradients in \eqref{eqn:cmp} with respect to the analog and digital precoders  can use the gradient formulations derived for \ac{pga} in \eqref{eqn:gr_Wa_k} and \eqref{eqn:gr_Wd_k}, respectively.
The gradient of $\emph{R} $ with respect to $\textbf{E}_b$ for each $b \in \mySet{B}$ is computed as (see  Appendix \ref{sec:dev Eb grad})
	\begin{align}
	\label{eqn:gr_Eb_k}
	\frac{\partial}{\partial\textbf{E}_b}
	\emph{R}(\textbf{W}_a, \{\textbf{W}_{d,b}\}, \{\textbf{H}_b \!+\!\textbf{E}_b\}) = \frac{1}{B}\textbf{G}_b(\textbf{W}_a ,\textbf{W}_{d,b} , \textbf{H}_b \!+ \!\textbf{E}_b)^{-T}(\tilde{\textbf{H}}_b^* \!+\! \textbf{E}_b^*) \textbf{W}_a^*\textbf{W}_{d,b}^*\textbf{W}_{d,b}^T\textbf{W}_a^T.
	\end{align}
While the above \ac{cmp} procedure introduces an additional iterative procedure which has to be carried out at each iteration of index $k$, it is typically sufficient to only carry out two iterations of \eqref{eqn:cmp} \cite{mokhtari2020unified}, i.e., repeat \eqref{eqn:cmp} for $i=1,\ldots,i_{\max}$ with $i_{\max}=2$.


{\bf Projection}: After the \ac{cmp} is conducted for two iterations,  the resulting matrices, denoted $\Big(\hat{\textbf{W}}_{a,(i_{\max})},$ $\{\hat{\textbf{W}}_{d,b,(i_{\max})}\}_{b\in \mySet{B}}, \{\hat{\textbf{E}}_{b,(i_{\max})}\}_{b\in \mySet{B}}\Big)$, are projected to meet the constraints. Thus, the update rule at iteration $k+1$ is
\begin{subequations}
\label{eqn:proj}
	\begin{align}
            \textbf{W}_a^{(k+1)} &= \Pi_{\mySet{A}} \left\{\hat{\textbf{W}}_{a,(i_{\max})}\right\}, \\
            \textbf{W}_{d,b}^{(k+1)} &= \sqrt{\frac{NB}{\sum_{b=1}^{B}\| \textbf{W}_a^{(k+1)}\hat{\textbf{W}}_{d,b,(i_{\max})} \|^2_F}} \cdot \hat{\textbf{W}}_{d,b,(i_{\max})}, \quad \forall b\in \mySet{B,} \\
            \textbf{E}_b^{(k+1)} &= \min\left\{ \frac{\varepsilon\cdot NM}{\| \hat{\textbf{E}}_{b,(i_{\max})} \|_F}, 1 \right\} \cdot \hat{\textbf{E}}_{b,(i_{\max})}, \quad \forall b\in \mySet{B},
	\end{align}
	\end{subequations}
where $\varepsilon$ is the error bound on the entries of the sub-channel matrix, $\{\textbf{H}_b\}_{b\in \mySet{B}} \in \mathbb{C}^{N\times{M}}$.

The overall procedure is summarized as Algorithm~\ref{alg:PCMP}. The matrices $\textbf{W}_a^{(0)}, \{\textbf{W}_{d,b}^{(0)}\}_{b\in \mySet{B}}$ are initialized in the same manner as in Algorithm~\ref{alg:PGA}, while $\{\textbf{E}_b^{(0)}\}_{b\in \mySet{B}}$ is generated randomly while normalizing to guarantee that $\|\textbf{E}_b\|_F<\epsilon$ for each $b \in \mySet{B}$. Notice that Algorithm~\ref{alg:PCMP} describes a gradient-based optimizer, similar to Algorithm~\ref{alg:PGA}. Therefore, its convergence speed depends as well on the step sizes $\mu_{a(i)}^{(k)}, \{\mu_{d,b(i)}^{(k)}\}_{b\in \mySet{B}}, \{\mu_{e,b(i)}^{(k)}\}_{b\in \mySet{B}}$, which are hard to select manually. In Subsection~\ref{subsec:rob l2o} we will describe the learn-to-rapidly-optimize method, extending the rationale used with full \ac{csi} in Algorithm~\ref{alg:L2O}, for accelerating Algorithm~\ref{alg:PCMP} convergence via data-aided hyperparameters tuning. 

  \begin{algorithm}
    \caption{Projected Conceptual Mirror Prox for Robust Hybrid Precoding}
    \label{alg:PCMP}
    \SetAlgoLined
    \SetKwInOut{Initialization}{Init}
    \Initialization{Randomize $\{\textbf{W}_{d,b}^{(0)}\}_{b\in \mySet{B}}, \{\textbf{E}_b^{(0)}\}_{b\in \mySet{B}}$ \newline
     $\textbf{W}_a^{(0)} \leftarrow$ first  $L$ right-singular vectors of $\frac{1}{B} \sum_b\tilde{\textbf{H}}_b$ \newline Set step sizes $\mu_{a(i)}^{(k)}, \{\mu_{d,b(i)}^{(k)}\}_{b\in \mySet{B}}, \{\mu_{e,b(i)}^{(k)}\}_{b\in \mySet{B}}$ }
    \SetKwInOut{Input}{Input}
    \Input{Channel matrices $\{\tilde{\textbf{H}}_b\}_{b\in \mySet{B}}$}  
    {
        \For{$k = 0, 1, \ldots$ until convergence}{
                    Set \eqref{eqn:init}
                    
                    \For{$i=1, \ldots, i_{\max}$}{
                    Calculate $\left( \hat{\textbf{W}}_{a,(i)}, 
                    \{\hat{\textbf{W}}_{d,b,(i)}\}_{b\in \mySet{B}}, \{\hat{\textbf{E}}_{b,(i)}\}_{b\in \mySet{B}} \right)$ by \eqref{eqn:cmp} \label{step:GradientsPCMP}
                    }
                    Update $\left( \textbf{W}_a^{(k+1)}, 
                    \{\textbf{W}_{d,b}^{(k+1)}\}_{b\in \mySet{B}}, \{\textbf{E}_b^{(k+1)}\}_{b\in \mySet{B}} \right)$ via \eqref{eqn:proj}  \label{stp:proj_pcmp}

                }
        \KwRet{$\{\textbf{W}_{d,b}^{(k)}\}_{b\in \mySet{B}}$ and $\textbf{W}_a^{(k)}$}
  }
\end{algorithm}

\subsection{Robust Learn-to-Optimize Hybrid Precoding}
\label{subsec:rob l2o}
Algorithm~\ref{alg:PCMP} optimizes hybrid precoders for a given noisy channel realization. Nevertheless, it needs to be executed in real-time, whenever there is a change in the channel, and thus the need to obtain reliable hybrid precoders rapidly, i.e., within a fixed and small number of iterations, still applies. To accomplish this, we use the deep unfolding methodology following the approach as in Subsection~\ref{subsec:SymbolDetDeep}, leveraging its ability to tune hyperparameters in optimization involving projected gradients of the rate function,  where the main differences  are in the number of learned parameters and in the algorithm structure. 

We  use the \ac{pcmp} method as the optimizer with exactly $\Kiter$ iterations and $i_{\max}=2$ internal iterations. Let us define the  step sizes vector for the analog update of the $k$th iteration as $\bm{\mu}_a^k \triangleq  (\mu_{a(1)}^{(k)} ,\mu_{a(2)}^{(k)})^T$; step sizes vectors for the digital updates of the $k$th iteration as $\{\bm{\mu}_{d,b}^k\}_{b\in \mySet{B}} \triangleq  \{(\mu_{d,b(1)}^{(k)} ,\mu_{d,b(2)}^{(k)})^T\}_{b\in \mySet{B}}$; and the step sizes vector for the error updates as  $\{\bm{\mu}_{e,b}^k\}_{b\in \mySet{B}} \triangleq  \{(\mu_{e,b(1)}^{(k)} ,\mu_{e,b(2)}^{(k)})^T\}_{b\in \mySet{B}}$. Accordingly, we define the $k$th iteration $2 \times 2B + 1$ step sizes matrix as $\bm{\mu}^k \triangleq (\bm{\mu}_a^k, \bm{\mu}_{d,1}^k, \ldots, \bm{\mu}_{d,B}^k, \bm{\mu}_{e,1}^k, \ldots, \bm{\mu}_{e,B}^k)$, obtaining the $\Kiter \times 2 \times 2B + 1$ step sizes tensor $\bm{\mu} \triangleq (\bm{\mu}^{0},  \ldots, \bm{\mu}^{\Kiter-1})$ for $\Kiter$ iterations of Algorithm~\ref{alg:PCMP}.

The unfolded architecture converts the \ac{pcmp} method with $K$ iteration into a discriminative algorithm whose trainable parameters are the entries of $\bm{\mu}$. Similarly to the approach used in Subsection~\ref{subsec:SymbolDetDeep}, we train the unfolded \ac{pcmp}  in an unsupervised manner, i.e., the data set only includes channel realizations, while  the level of tolerable error $\varepsilon$ is given.
The loss function for a given normalized channel $\tilde{\textbf{H}} = \{\tilde{\textbf{H}}_b\}_{b\in \mySet{B}}$ and step sizes $\bm{\mu}$ is computed as the maximal negative resulting achievable sum-rate of this channel, \eqref{eqn:achievable rate} when the maximum is taken over all $\| \textbf{E}_b \|_F < \varepsilon$, and the precoders are $\Big(\textbf{W}_a^{(\Kiter)}, \{\textbf{W}_{d,b}^{(\Kiter)}\}_{b\in \mySet{B}}\Big) \triangleq \textbf{PCMP}_{\Kiter}(\tilde{\textbf{H}} , \bm{\mu}) $, i.e., the precoders obtained via Algorithm~\ref{alg:PCMP} with $\Kiter$ iterations and step sizes $\bm{\mu}$. The resulting loss is
\begin{align}
\label{eqn:Maxloss}
    \!\! \mathcal{L}(\tilde{\textbf{H}}, \bm{\mu}) \! = \! \max_{\| \textbf{E}_b \|_F < \varepsilon} \!\left\{ \frac{-1}{B}  \sum_{b=1}^{B}  \log \left|\textbf{I}_N \!  +\! \frac{1}{N\sigma^2} ({\textbf{H}}_b \! + \!\textbf{E}_b) \textbf{W}_a^{(K)}\textbf{W}_{d,b}^{(K)} \big(({\textbf{H}}_b \!+\! \textbf{E}_b)\textbf{W}_a^{(K)}\textbf{W}_{d,b}^{(K)}\big)^H\right| \right\}.
\end{align}
The maximization term in \eqref{eqn:Maxloss} notably complicates its usage as a training objective for learning the \ac{pcmp} hyperparameters. To overcome this, we generate a {\em finite} set of $n_e$  error patterns (satisfying $\| \textbf{E}_b \|_F < \varepsilon$)   via random generation, to which we add the zero error, i.e., $\mySet{E} \triangleq \bigl \{ \{\textbf{E}_b^t\}_{b\in\mySet{B}} \bigl \}_{t=1}^{n_e} \cup \{\myMat{0}\}$, and seek the setting which minimizes the maximal loss among these error patterns. Namely, to maintain feasible training, the objective used for training the unfolded algorithm evaluates the loss based on its output after $K$ iterations and is given by  
\begin{align}
\label{eqn:Maxloss2}
   \!\!  \tilde{\mathcal{L}}(\tilde{\textbf{H}}, \bm{\mu}) \! =\!  \max_{ \textbf{E}_b  \in \mySet{E} } \!\left\{ \frac{-1}{B}  \sum_{b=1}^{B}  \log \left|\textbf{I}_N   \!+\! \frac{1}{N\sigma^2} ({\textbf{H}}_b \!+\! \textbf{E}_b) \textbf{W}_a^{(K)}\textbf{W}_{d,b}^{(K)} \big(({\textbf{H}}_b \!+\! \textbf{E}_b)\textbf{W}_a^{(K)}\textbf{W}_{d,b}^{(K)}\big)^H\right| \right\}.
\end{align}


The robust learn-to-optimize method, summarized as Algorithm~\ref{alg:L2O_rob}, uses data to tune $\bm{\mu}$.
We initialize $\bm{\mu}$ before the training process, with fixed step sizes with which \ac{pcmp} converges (though not necessarily rapidly). 
After training, the learned matrix $\bm{\mu}$ is used as hyperparameters for rapidly converting a noisy channel realization into a hybrid precoding setting via $\Kiter$ iterations of Algorithm~\ref{alg:PCMP}.

  \begin{algorithm}
    \caption{Robust Learn-to-Optimize Hybrid Precoding}
    \label{alg:L2O_rob}
    \SetAlgoLined
    \SetKwInOut{Initialization}{Init}
    \Initialization{Set $\bm{\mu}$ as fixed step sizes. Fix learning rate $\eta$}
    \SetKwInOut{Input}{Input}
    \SetKwInOut{Output}{Output}
    \SetKwFor{RepTimes}{repeat}{times}{end}
    \Input{
    Training set  $\mathcal{D} = \{\tilde{\textbf{H}}^r\}_{r=1}^{|\mathcal{D}|}$}  
    {
        \For{${\rm epoch} = 0, 1, \ldots, {\rm epoch}_{\max}-1$}{%
                    Randomly divide  $\mathcal{D}$ into $Q$ batches $\{\mathcal{D}_q\}_{q=1}^Q$
                    
                    \For{$q = 1, \ldots, Q$}{
                    
                    Compute precoders via $\textbf{PCMP}_{\Kiter}(\mathcal{D}_q , \bm{\mu}) $
                    
                    Compute the average loss of the batch: $\mathcal{L}(\bm{\mu}) = \frac{1}{|\mathcal{D}_q|} \sum_{\tilde{\textbf{H}} \in \mathcal{D}_q}\tilde{\mathcal{L}}(\tilde{\textbf{H}}, \bm{\mu})$
                    
                    Update  $\bm{\mu}\leftarrow \bm{\mu} - \eta\nabla_{\bm{\mu}}\mathcal{L}(\bm{\mu})$ \label{stp:updatee1}
                    }
                    
                    }
        \KwRet{$\bm{\mu}$}
  }
\end{algorithm}

\subsection{Discussion}
\label{subsec:Discussion2}
The proposed robust learn-to-optimize method in Algorithm~\ref{alg:L2O_rob} extends the method in Algorithm~\ref{alg:L2O} to cope with noisy \ac{csi}. In particular, for the special case of $\varepsilon=0$ and $i_{\max} =1$, it holds that the \ac{pcmp} algorithm (Algorithm~\ref{alg:PCMP}) reduces to the \ac{pga} method (Algorithm~\ref{alg:PGA}). Accordingly, the data-aided optimized Algorithm~\ref{alg:L2O_rob}  specializes Algorithm~\ref{alg:L2O}. Consequently, our gradual design, which started with designing learned \ac{pga} optimization applied to the rate function in  Section~\ref{sec:Hybrid}, allowed us to extend its derivation to realize robust learned \ac{pcmp} algorithm for maximin rate optimization. In addition, this implementation shows that the general learn-to-optimize method can be adapted for speeding the convergence of a broad range iterative model-based hyperparameters-depending algorithms, and particularly those utilizing gradient and projection steps.   

In terms of complexity, the unfolded \ac{pcmp} with $K$ iterations and $i_{\max} = 2$ shares the same complexity order as that the \ac{pga} method with full \ac{csi}. In particular, 
 Step~\ref{step:GradientsPCMP} of Algorithm~\ref{alg:PCMP} repeats the gradient computations of Algorithm~\ref{alg:PGA} $i_{\max}$ times per iteration and includes additional gradients with respect to $\textbf{E}_b$ (which is of the same complexity as taking the gradients with respect to $\textbf{W}_{d,b}$). Thus, the complexity order of the unfolded \ac{pcmp} algorithm is 
 \begin{align}
 \label{eqn:CompPCMP}
      \mySet{C}_{\rm PCMP} &= \mySet{O}\Bigl(\Kiter \cdot B M \bigl(  N  L + M (N+L)i_{\max} \bigl) \Bigl),
 \end{align}
 which, for  $i_{\max} = 2$, yields a similar complexity as that of the unfolded \ac{pga} in \eqref{eqn:CompPGA}.

The ability to rapidly optimize in a robust manner via the unfolded \ac{pcmp} results in hybrid precoders that are
less sensitive to channel estimation errors. 
The \ac{pcmp} algorithm is coping with mismatched channels by considering a wide range of channel errors when setting the hybrid precoders. While the \ac{pga} algorithm is carried out each time the channel changes, the \ac{pcmp} algorithm can be carried out less frequently since it is more robust and stable. Accordingly, the \ac{pcmp} can be also used to reduce the rate in which the optimization procedure is carried out, and not only for coping with mismatched channels.

The error bound value, $\varepsilon$, plays a key role in the robust algorithm. It affects the robustness, reliability, and stability of the solution, and therefore, it needs to be chosen based on some prior knowledge, or on application needs. Another approach is to use channel estimation methods prior to the precoders design we presented in this work, and by this overcome the mismatched channel problem. An additional important parameter is the setting of the expected error patterns evaluated during training, i.e., $\mySet{E}$. While in our evaluation we generated this set in a random fashion and included in it the zero-error pattern to guarantee suitability for error-free case, one can consider alternative methods for preparing this set such that the resulting optimized hyperparameters would be most suitable for the desired robust optimization metric. Since this setting is carried out offline and thus complexity is not a key issue here, one can possibly explore the usage of sequential sampling and Bayesian optimization techniques \cite{frazier2018tutorial} for setting  $\mySet{E}$. 
We leave the study of extensions of our method to future work.





\section{Numerical Evaluations}
\label{sec:Sims}
We next numerically evaluate the proposed unfolded optimization framework\footnote{The complete source is available at \url{https://github.com/ortalagiv/Learn-to-Rapidly-Optimize-Hybrid-Precoding}.}. Our main purpose is to numerically  demonstrate the ability of the proposed learn-to-optimize framework to notably reduce the number of iterations compared with conventional optimization with fixed hyperparameters. 

Our numerical evaluations commence with settings with full \ac{csi} in Subsection~\ref{subsec:pga}. As the full \ac{csi} setting (with analog combiners constrained to represent phase shifter network) is the common setup considered for hybrid precoding, this numerical evaluation allows us to compare the first step of our design, i.e., the unfolded \ac{pga} method, to existing optimization methods. Then,   we consider robust optimization based on the unfolded \ac{pcmp} in Subsection~\ref{subsec:pcmp}. For both \ac{csi} settings, we simulate two different models for the downlink hybrid \ac{mimo} system as described in Subsection~\ref{subsec:MIMO}: Rayleigh fading, where the channel matrix in each frequency bin is randomized from an i.i.d. Gaussian distribution; and the \ac{quadriga} model \cite{jaeckel2014quadriga}, an open source geometry-based stochastic channel model, relevant for simulating realistic \ac{mimo} systems.


\subsection{Hybrid Precoding with Full \ac{csi}}
\label{subsec:pga}
For full \ac{csi}, the optimizer for designing the hybrid precoder is \ac{pga}, and its data-aided learn-to-optimize version is detailed Subsection~\ref{subsec:SymbolDetDeep}. The implementation of this method can be carried out for both unconstrained analog combiners (for which the only constraint is the power constraint), as well as hardware-limited designs, where a common constraint is that of phase shifter networks. We thus divide our evaluation into learned \ac{pga} with unconstrained analog combiners and \ac{pga} with phase shifters. 

\subsubsection{\ac{pga} using unconstrained analog combiners}
\label{subsubsec:uc_pga}
In this case we use the analog projection operator as $\Pi_{\mySet{A}}(\textbf{A})=\textbf{A}$, i.e., we do not impose any constraints on the analog architecture. Such unconstrained combiners can be approximated using architectures proposed in, e.g.~\cite{gong2019rf,zirtiloglu2022power}.

We consider three different settings for the number frequency bands $B$, the number of users $N$, the number of RF chains $L$, the number of transmit antennas $M$, and for the data source: \\
$1)$ A Rayleigh channel with $8$ frequency bins, configured with 
$B=8, N=6, L=10, M=12$; \\
$2)$ A \ac{quadriga} channel with $B=16$ frequency bins, where we use  $ N=4, L=6, M=12$; \\ $3)$ A wideband \ac{quadriga} channel with $B=128, N=4, L=8, M=16$.  \\
For each configuration, we applied Algorithm~\ref{alg:L2O} to learn to set hybrid precoders with merely $K=5, 10$ iterations based on $|\mySet{D}|=1000$ channel realization, where we used $50-70$ epochs with  batch size $|\mySet{D}_q|=100$ and used Adam for the update in Step~\ref{stp:update1} of Algorithm~\ref{alg:L2O}. 

\begin{figure}
\begin{center}
%
%
\begin{subfigure}[pt]{0.49\linewidth}
\includegraphics[width=\linewidth]{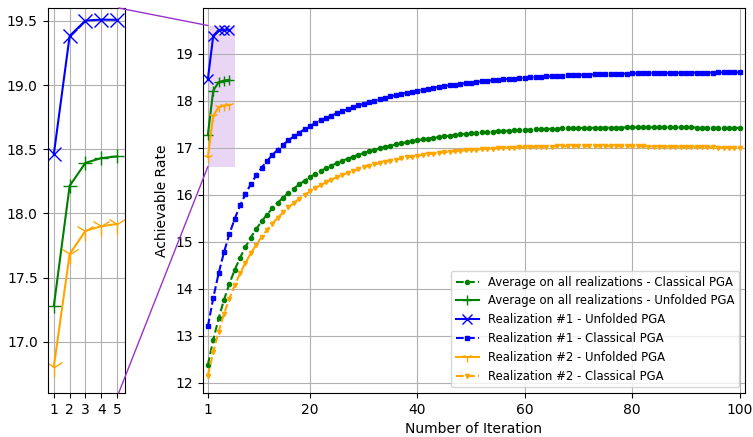}
\caption{Sum-rate per \ac{pga} iteration.}
\label{fig:I1a}
\end{subfigure}
%
%
\begin{subfigure}[pt]{0.49\linewidth} 
\includegraphics[width=\columnwidth]{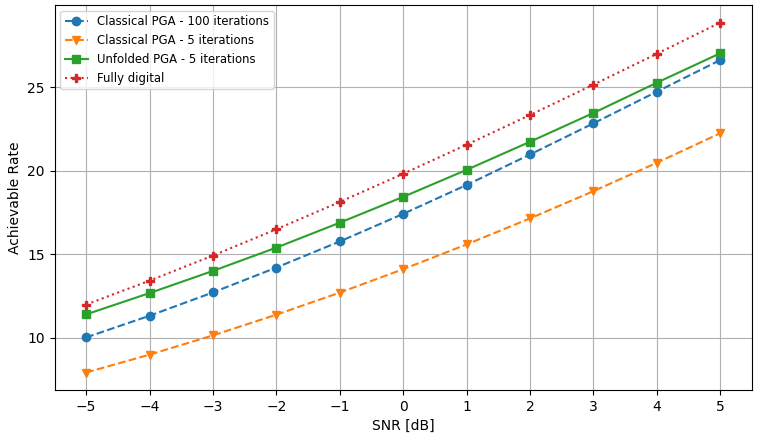}
\caption{Sum-rate vs. \ac{snr}.}
\label{fig:I1b}
\end{subfigure} 
\caption{Rayleigh channel; unconstrained analog precoder; $B=8, N=6, L=10, M=12$.}
\label{fig:I1}
\end{center} 
\end{figure}

\begin{figure}
\begin{center}
%
%
\begin{subfigure}[pt]{0.49\linewidth}
\includegraphics[width=\linewidth]{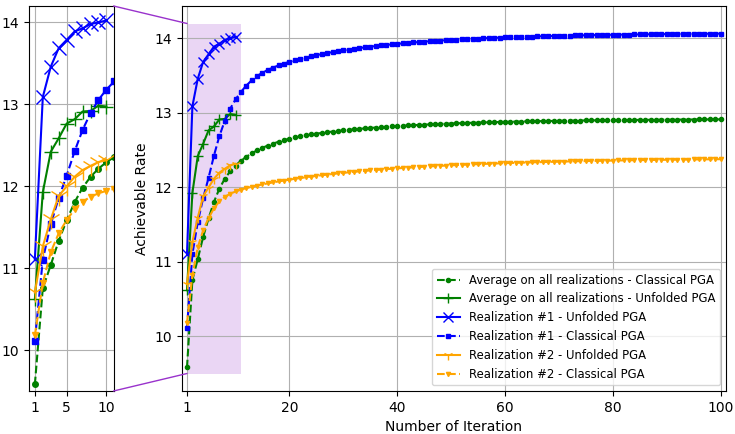}
\caption{Sum-rate per \ac{pga} iteration.}
\label{fig:I2a}
\end{subfigure}
%
%
\begin{subfigure}[pt]{0.49\linewidth} 
\includegraphics[width=\columnwidth]{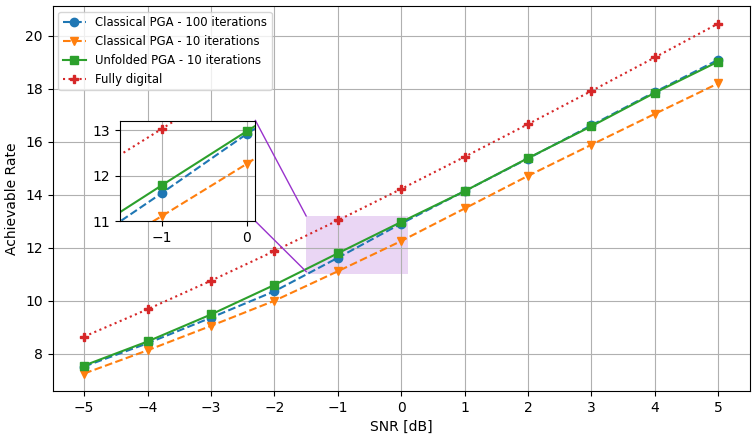}
\caption{Sum-rate vs. \ac{snr}.}
\label{fig:I2b}
\end{subfigure} 
\caption{\ac{quadriga} channel; unconstrained analog precoder; $B=16, N=4, L=6, M=12$.}
\label{fig:I2}
\end{center} 
\end{figure}

\begin{figure}
\begin{center}
%
%
\begin{subfigure}[pt]{0.49\linewidth}
\includegraphics[width=\linewidth]{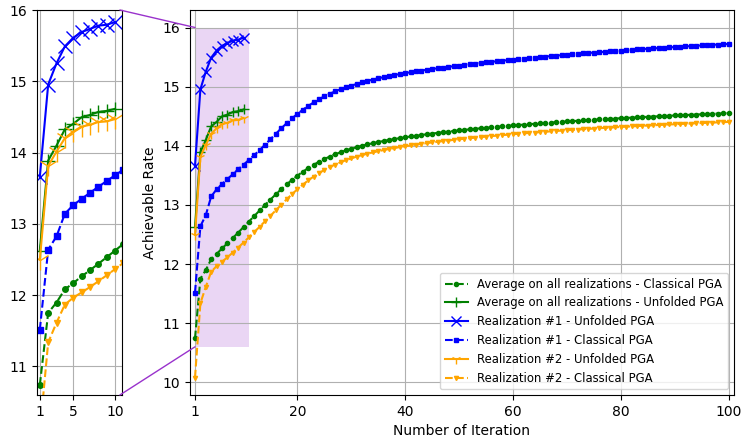}
\caption{Sum-rate per \ac{pga} iteration.}
\label{fig:I3a}
\end{subfigure}
%
%
\begin{subfigure}[pt]{0.49\linewidth} 
\includegraphics[width=\columnwidth]{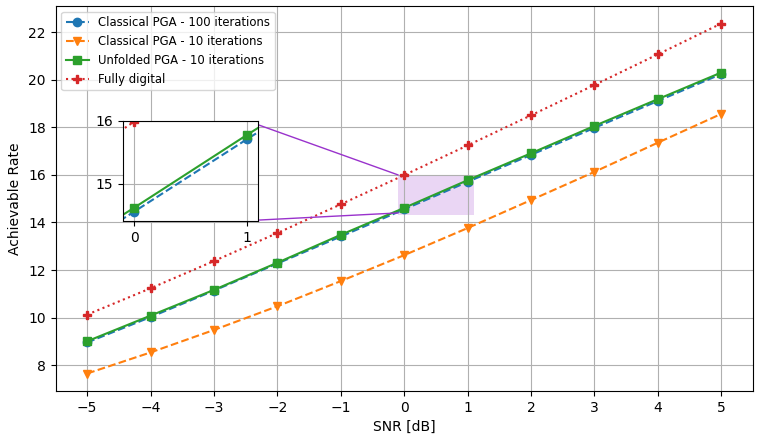}
\caption{Sum-rate vs. \ac{snr}.}
\label{fig:I3b}
\end{subfigure} 
\caption{\ac{quadriga} channel; unconstrained analog precoder; $B=128, N=4, L=8, M=16$.}
\label{fig:I3}
\end{center} 
\end{figure}

The \ac{pga} hybrid precoding design with $K$ optimized iterations is compared with applying \ac{pga} with fixed hyperparameters, where we used a constant step size chosen based on empirical trials. Both hybrid precoders are evaluated over $100$ unseen test channels. The simulation results for the settings of $(B=8, N=6, L=10, M=12)$, $(B=16, N=4, L=6, M=12)$, and $(B=128, N=4, L=8, M=16)$ are depicted in Fig.~\ref{fig:I1}, Fig.~\ref{fig:I2}, and Fig.~\ref{fig:I3}, respectively, where each figure compares both the convergence of the algorithms averaged over all channel realizations and for the randomly chose realizations, as well as the resulting sum-rates versus \ac{snr}.

The resulting sum-rates (for \ac{snr}$=0$dB) versus the number of \ac{pga} iterations, of both the unfolded \ac{pga} and the classical \ac{pga} (with manually chosen constant step sizes), when averaged over all unseen channels, and for two random channel realizations, are depicted in Fig.~\ref{fig:I1a}, Fig.~\ref{fig:I2a}, and Fig.~\ref{fig:I3a}. We observe in these figures that the proposed learn-to-optimize consistently facilitates simple \ac{pga} optimization of the hybrid precoders, achieving similar and even surpassing the sum-rates achieved by conventional \ac{pga} with fixed step sizes, while requiring much fewer iterations. The reduction in the number of iterations due to learn-to-optimize is by factors of $20$ ($5$ iterations vs. $100$ iterations) and $10$ ($10$ iterations vs. $100$ iterations) in speed, compared to conventional fixed-size optimization. These gains are consistent over the different channel models and configurations considered. 
Fig.~\ref{fig:I1b}, Fig.~\ref{fig:I2b}, and Fig.~\ref{fig:I3b} illustrate the sum-rate for different values of \ac{snr}, of the proposed unfolded \ac{pga} compared with that of the classical \ac{pga} and the sum-rate resulting from fully digital baseband precoding, serving as an upper bound on the achievable sum-rate. The figures show the gain of converting the classical \ac{pga} algorithm in a discriminative trainable model. In particular, the unfolded algorithm implements \ac{pga} with merely few iterations while systematically improving upon conventional \ac{pga} with fixed step size and $100$ iterations. 
These results demonstrate the benefits of the proposed approach in leveraging data to improve both performance and convergence speed while preserving the interpretability and suitability of conventional iterative optimizers.  

%
%

%
%

%
%

\subsubsection{\ac{pga} using phase shifters networks}
\label{subsubsec:c_pga}
The numerical evaluation above showed that the considered deep unfolding methodology indeed facilitates rapid optimization of hybrid precoders. To demonstrate that these gains are not unique to unconstrained analog combiners, as well as to compare with alternative iterative optimizers, we next consider phase shifters networks for the implementation of the analog architecture, as often considered in the literature. This means that the projection in \eqref{eqn:pr_Wa_partially} is used in Step~\ref{line:P_Wa} of Algorithm~\ref{alg:PGA}. We used as a benchmark the MO-AltMin algorithm of \cite{yu2016alternating}, which is designed to iteratively optimize such constrained hybrid precoders.

\begin{figure}
\begin{center}
%
%
\begin{subfigure}[pt]{0.49\linewidth}
\includegraphics[width=\linewidth]{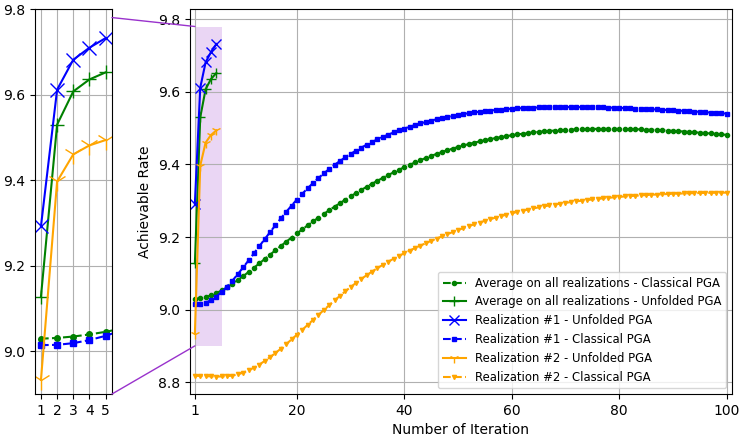}
\caption{Sum-rate per \ac{pga} iteration.}
\label{fig:I4a}
\end{subfigure}
%
%
\begin{subfigure}[pt]{0.49\linewidth} 
\includegraphics[width=\columnwidth]{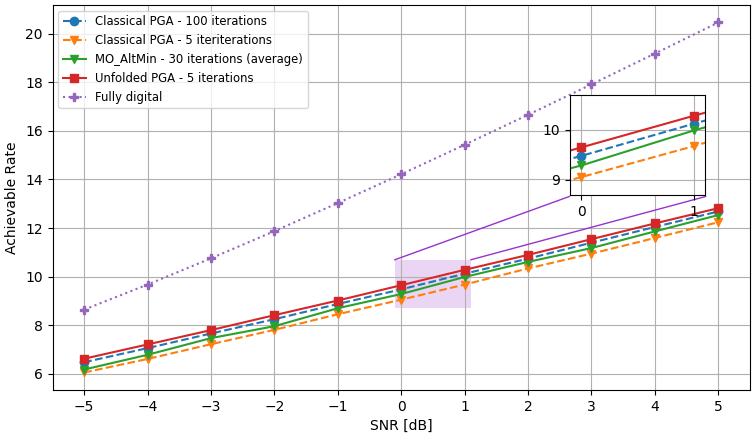}
\caption{Sum-rate vs. \ac{snr}.}
\label{fig:I4b}
\end{subfigure} 
\caption{\ac{quadriga} channel; constrained analog precoder; $B=16, N=4, L=2, M=12$.}
\label{fig:I4}
\end{center} 
\end{figure}

\begin{figure}
\begin{center}
%
%
\begin{subfigure}[pt]{0.49\linewidth}
\includegraphics[width=\linewidth]{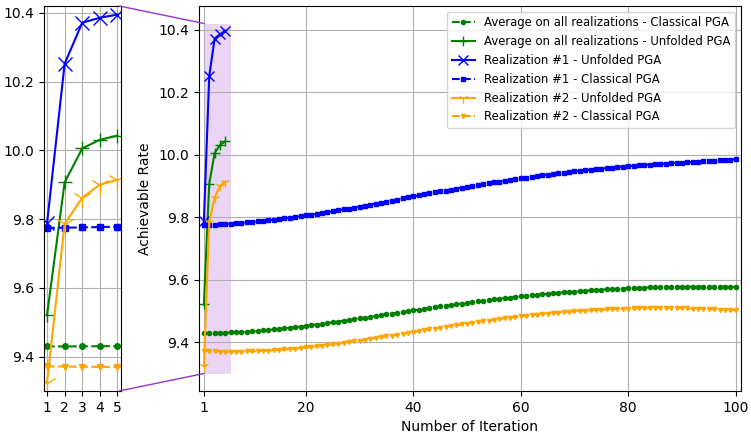}
\caption{Sum-rate per \ac{pga} iteration.}
\label{fig:I5a}
\end{subfigure}
%
%
\begin{subfigure}[pt]{0.49\linewidth} 
\includegraphics[width=\columnwidth]{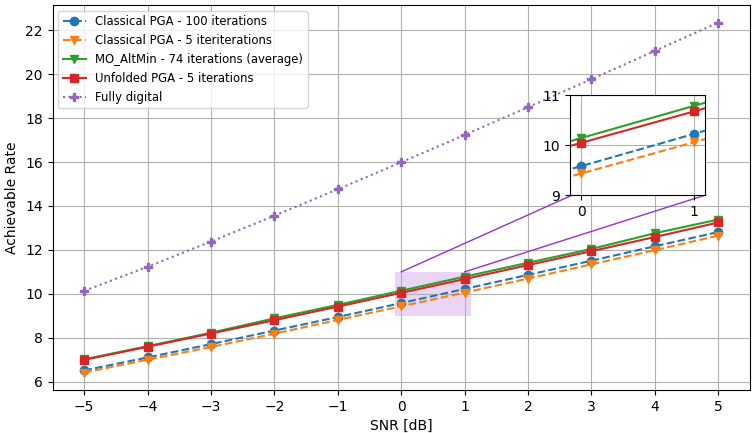}
\caption{Sum-rate vs. \ac{snr}.}
\label{fig:I5b}
\end{subfigure} 
\caption{\ac{quadriga} channel; constrained analog precoder; $B=128, N=4, L=2, M=16$.}
\label{fig:I5}
\end{center} 
\end{figure}

We consider the \ac{quadriga} channel model for two different settings of the number frequency bands $B$, users $N$, RF chains $L$, and transmit antennas$M$: $1)$ $B=16, N=4, L=2, M=12$; and $2)$ $B=128, N=4, L=2, M=16$.
Similarly to the unconstrained case, for each setting, we apply Algorithm~\ref{alg:L2O} to learn the hyperparameters for which the \ac{pga} converges with $K=5$ iterations. In the training procedure we use $|\mySet{D}|=1000$ channels for each setting, where we used $50$ epochs with batch size $|\mySet{D}_q|=100$ and used Adam for the update in Step~\ref{stp:update1}. The \ac{pga} with $K=5$ learned iterations is compared with the \ac{pga} with fixed hyperparameters, chosen empirically. The hybrid precoders optimizers are evaluated over $100$ unseen test channels. 
The performance evaluation results are shown in Fig.~\ref{fig:I4}, and Fig.~\ref{fig:I5}, for the settings of $B=16, N=4, L=2, M=12$, and $B=128, N=4, L=2, M=16$, respectively, where each figure examines both the convergence curves as well as the sum-rate achieved at the end of the optimization procedure versus \ac{snr}.

The resulting sum-rate (for \ac{snr}$=0$dB) versus the number of \ac{pga} iterations, when averaged over all unseen channels, and of two random channel realizations, is shown in Fig.~\ref{fig:I4a}, and Fig.~\ref{fig:I5a}. We observe in these figures that the learn-to-optimize method is able to accelerate and outperform the \ac{pga} optimization, when compared to the standard \ac{pga}, with constant, manually chosen, step sizes. Observe that the number of iterations is reduced by $20$ ($5$ iterations vs. $100$), compared to standard optimization. The gain in performance systematically observed here follows from the ability of data-driven optimization to facilitate coping with the non-convex nature of the resulting optimization problem.
Fig.~\ref{fig:I4b}, and Fig.~\ref{fig:I5b} demonstrate the comparison between the proposed unfolded \ac{pga}, the classical \ac{pga}, the MO-Altmin algorithm of~\cite{yu2016alternating}, and  fully digital baseband precoding. It is shown that the gap from the fully digital baseband precoding is significant, due to the use of a relatively small number of RF chains in the hybrid architecture. When comparing the iterative algorithms we see a small difference in terms of sum-rate, but it is important to consider the fact that the proposed unfolded \ac{pga} operates with the least number of iterations, therefor it achieves improvement in terms of speed when comparing to the benchmark and to the standard \ac{pga}.
These results show the ability of the proposed learn-to-optimize technique to tackle and incorporate different constraints into the \ac{pga} algorithm while enabling rapid performance and fully preserving interpretability. 

%
%

%
%

\subsection{Hybrid Precoding with Noisy \ac{csi}}
\label{subsec:pcmp}
The numerical studies reported in Subsection~\ref{subsec:pga} empirically validate the ability of the proposed methodology to facilitate rapid optimization under different settings. We next demonstrate its ability to simultaneously enable rapid and robust hybrid precoding. Hence, in the following, we consider the unfolding of the \ac{pcmp} algorithm for precoding under noisy channel estimation. Having demonstrated the ability of our approach to deal with constrained precoders, we focus here on unconstrained analog combiners. According, we implement of the learn-to-optimize methodology, i.e. Algorithm~\ref{alg:L2O_rob}, for Algorithm~\ref{alg:PCMP}, when the projection operator applied on the analog precoder in Step~\ref{stp:proj_pcmp} is set to be $\Pi_{\mySet{A}}(\textbf{A})=\textbf{A}$. The set of considered error patterns $\mySet{E}$ is comprised of $n_e = 20$ patterns randomly generated with Frobenius norm values in the range  $(0,\epsilon)$, and we evaluate the maximin rate over $\mySet{E}$.

\begin{figure}
\begin{center}
%
%
\begin{subfigure}[pt]{0.49\linewidth}
\includegraphics[width=\linewidth]{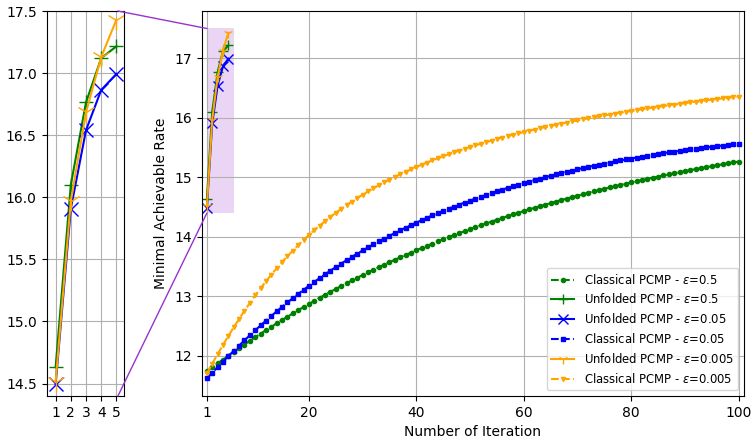}
\caption{Minimal sum-rate per \ac{pcmp} iteration.}
\label{fig:I6a}
\end{subfigure}
%
%
\begin{subfigure}[pt]{0.49\linewidth} 
\includegraphics[width=\columnwidth]{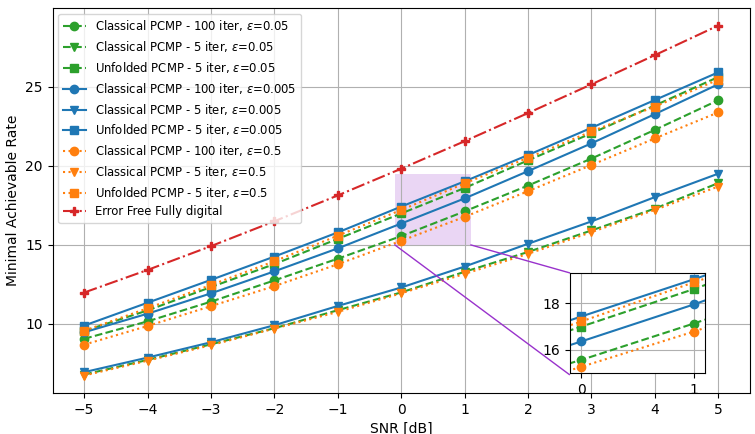}
\caption{Minimal sum-rate vs. \ac{snr}.}
\label{fig:I6b}
\end{subfigure} 
\caption{Rayleigh channel; unconstrained analog precoder; $B=8, N=6, L=10, M=12$.}
\label{fig:I6}
\end{center} 
\end{figure}

\begin{figure}
\begin{center}
%
%
\begin{subfigure}[pt]{0.49\linewidth}
\includegraphics[width=\linewidth]{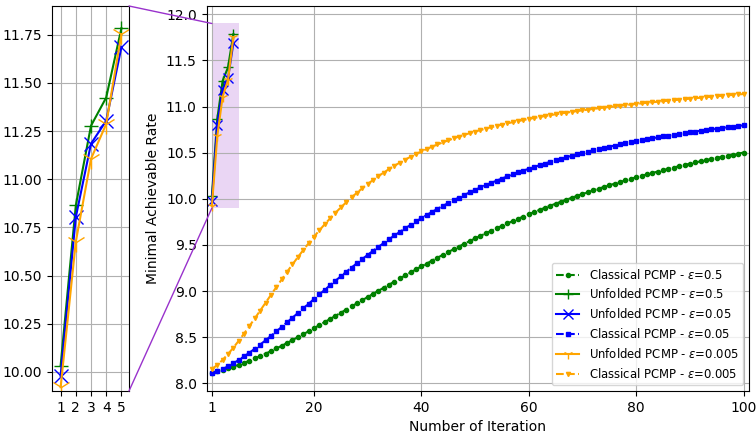}
\caption{Minimal sum-rate per \ac{pcmp} iteration.}
\label{fig:I7a}
\end{subfigure}
%
%
\begin{subfigure}[pt]{0.49\linewidth} 
\includegraphics[width=\columnwidth]{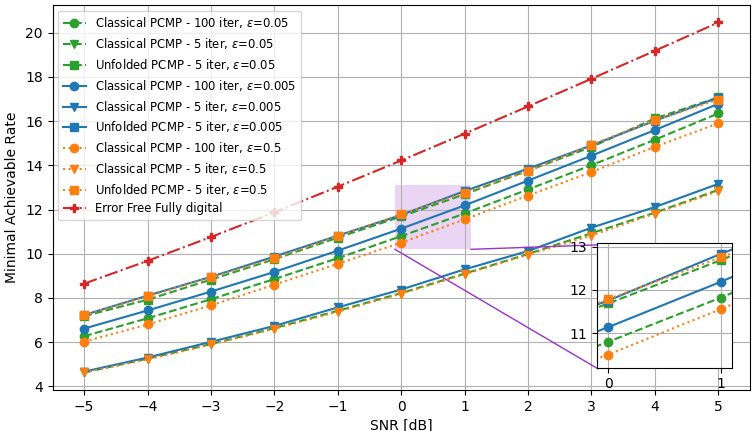}
\caption{Minimal sum-rate vs. \ac{snr}.}
\label{fig:I7b}
\end{subfigure} 
\caption{\ac{quadriga} channel; unconstrained analog precoder; $B=16, N=4, L=6, M=12$.}
\label{fig:I7}
\end{center} 
\end{figure}

We examine two different settings of the number frequency bands $B$, users $N$, RF chains $L$, transmit antennas $M$, and the data source: $B=8, N=6, L=10, M=12$, with Rayleigh channels; and $B=16, N=4, L=6, M=12$ with \ac{quadriga} channel model, the simulations results for each setting are demonstrated in Fig.~\ref{fig:I6}, and Fig.~\ref{fig:I7}, respectively. Again, we evaluate both the convergence rate and the minimal rate (within the tolerable error regime) at the end of the optimization procedure versus  \ac{snr}, and compare our results with the conventional \ac{pcmp} optimizer. In the simulations, three error bound values are considered, $\varepsilon = 0.005, 0.05, 0.5$. For each setting, we applied Algorithm~\ref{alg:L2O_rob} to learn to set hybrid precoders with $K=5$ iterations based on $|\mySet{D}|=1000$ channels, where we used $40-50$ epochs with batch size $|\mySet{D}_q|=100$, and used Adam for the update in Step~\ref{stp:updatee1}. We compared the standard \ac{pcmp}, with constant hyperparameters, to the optimized \ac{pcmp} with exactly $\Kiter = 5$ iterations.

Fig.~\ref{fig:I6a} and Fig.~\ref{fig:I7a} depict the minimal sum-rates, averaged on $100$ unseen test channels results, for the three error bound values, vs. the number of \ac{pcmp} iterations. In these figures, we observe notable gains in convergence speed of a factor of $20$ ($5$ iterations vs. $100$ iterations), and it is shown that the performances of the learned \ac{pcmp} algorithm consistently surpasses the performances of the conventional \ac{pcmp} with constant step sizes. 
In Fig.~\ref{fig:I6b}, and Fig.~\ref{fig:I7b}, the sum-rate for different values of \ac{snr} is shown. These figures include the fully digital baseband precoding error-free sum-rates, which are calculated with full \ac{csi}, and reflect on the best performance one can achieve given full \ac{csi} and without  RF chain reduction.
We observe that the proposed robust optimization method, the unfolded \ac{pcmp}, achieves relatively close performances to the full-\ac{csi} fully digital baseband precoding, systematically outperforming the model-based \ac{pcmp} operating with much more iterations. These results demonstrates the gains and advantages of our proposed learn-to-optimize method in enabling optimization of hybrid precoders which is both rapid and robust.


%
%

%
%

\section{Conclusions}
\label{sec:Conclusions}
In this work we proposed a method to leverage data to enable rapid, robust, and interpretable tuning of hybrid precoders. Our approach unfolds a suitable optimizer for maximizing the minimal sum-rate within a given tolerable \ac{csi} error into a fixed and small number of iterations. Then, we use data to tune the hyperparameters of each iteration.
Our method is  shown to notably improve convergence speed while setting hybrid precoders which achieve similar and even improved sum-rates compared to those tuned via lengthy non-learned optimization.

\section{Acknowledgements}
\label{sec:Acknowledgements}
The authors would like to thank Tomer Yeblonka from CEVA for his valuable inputs and meaningful discussions.
 
 \appendices
\section{Complex Gradients of $\emph{R}(\cdot)$ with Respect to $\textbf{W}_a, \{\textbf{W}_{d,b}\}_{b\in \mySet{B}}$}
\label{sec:dev Wa grad}
We first derive \eqref{eqn:gr_Wa_k} following the computation of complex-valued gradients of a general scalar function detailed in  \cite{hjorungnes2011complex}. 
The complex differential of $\emph{R}\left(\textbf{W}_a , \{\textbf{W}_{d,b}\}, \{\textbf{H}_b\}\right)$ with respect to $\textbf{W}_a$ is 
	\begin{align*}
	&d_{\textbf{W}_a}\emph{R}\left(\textbf{W}_a , \{\textbf{W}_{d,b}\}, \{\textbf{H}_b\}\right)  \notag
	\\&=d_{\textbf{W}_a} \left( \frac{1}{B} \sum_{b=1}^{B} \log \left|\textbf{I}_N + \tilde{\textbf{H}}_b\textbf{W}_a\textbf{W}_{d,b} \notag \textbf{W}_{d,b}^H\textbf{W}_a^H\tilde{\textbf{H}}_b^H\right| \right) 
	\\ &= {\rm Tr}\Big[\frac{1}{B} \sum_{b=1}^{B}\textbf{W}_{d,b}\textbf{W}_{d,b}^H\textbf{W}_a^H\tilde{\textbf{H}}_b^H\textbf{G}_b(\textbf{W}_a, \textbf{W}_{d,b}, \textbf{H}_b)^{-1}\tilde{\textbf{H}}_b(d\textbf{W}_a)\Big] \notag
	\\&\quad+ {\rm Tr}\Big[\frac{1}{B} \sum_{b=1}^{B}\textbf{W}_{d,b}^*\textbf{W}_{d,b}^T\textbf{W}_a^T\tilde{\textbf{H}}_b^T\textbf{G}_b(\textbf{W}_a, \textbf{W}_{d,b}, \textbf{H}_b)^{-1}\tilde{\textbf{H}}_b^*(d\textbf{W}_a^*)\Big]. 
	\end{align*}
Using \cite[Table 3.2]{hjorungnes2011complex}, this yields \eqref{eqn:gr_Wa_k}.

Similarly, the  differential of $\emph{R}\left(\textbf{W}_a , \{\textbf{W}_{d,b}\}, \{\textbf{H}_b\}\right)$ with respect to $\textbf{W}_{d,b}$ is 
	\begin{align*}
	&d_{\textbf{W}_{d,b}}\emph{R}\left(\textbf{W}_a , \{\textbf{W}_{d,b}\}, \{\textbf{H}_b\}\right)  \notag
	\\&=d_{\textbf{W}_{d,b}} \left( \frac{1}{B} \sum_{n=1}^{B} \log \left|\textbf{I}_N + \tilde{\textbf{H}}_n\textbf{W}_a\textbf{W}_{d,n} \textbf{W}_{d,n}^H\textbf{W}_a^H\tilde{\textbf{H}}_n^H\right| \right)  \notag
	\\ &= {\rm Tr}\Big[\frac{1}{B}\textbf{W}_{d,b}^H\textbf{W}_a^H\tilde{\textbf{H}}_b^H \tilde{\textbf{H}}_b\textbf{W}_a(d\textbf{W}_{d,b})\Big]  \notag
	\\ &\quad+ {\rm Tr}\Big[\frac{1}{B} \textbf{W}_{d,b}^T\textbf{W}_a^T\tilde{\textbf{H}}_b^T \textbf{G}_b(\textbf{W}_a, \textbf{W}_{d,b}, \textbf{H}_b)^{-T} \tilde{\textbf{H}}_b^*\textbf{W}_a^* (d\textbf{W}_{d,b}^*)\Big]. 
	\end{align*}
Using \cite[Table 3.2]{hjorungnes2011complex}, this yields \eqref{eqn:gr_Wd_k}.

\vspace{-0.2cm}
\section{Complex Gradient of $\emph{R}(\cdot)$ with Respect to $\{\textbf{E}_b\}_{b\in \mySet{B}}$}
\label{sec:dev Eb grad}

The complex differential of $\emph{R}(\textbf{W}_a, \{\textbf{W}_{d,b}\}, \{\textbf{H}_b +\textbf{E}_b\})$ with respect to $\textbf{E}_b$ is given by
	\begin{align*}
	&d_{\textbf{E}_b}\emph{R}(\textbf{W}_a, \{\textbf{W}_{d,b}\}, \{\textbf{H}_b +\textbf{E}_b\})\\ &= 
	d_{\textbf{E}_b}\left(\frac{1}{B} \sum_{n=1}^{B} \log \left|\textbf{I}_N \!+\!  \left(\tilde{\textbf{H}}_n + \textbf{E}_n\right)\textbf{W}_a\textbf{W}_{d,n} \textbf{W}_{d,n}^H\textbf{W}_a^H\left(\tilde{\textbf{H}}_n + \textbf{E}_n\right)^H\right|\right)\\
  &= {\rm Tr}\Big[ \frac{1}{B} \Big(  \textbf{W}_a\textbf{W}_{d,b}\textbf{W}_{d,b}^H\textbf{W}_a^H\tilde{\textbf{H}}_b^H \textbf{G}_b(\textbf{W}_a, \textbf{W}_{d,b}, \textbf{H}_b +\textbf{E}_b)^{-1} \\
  & \quad+ \textbf{W}_a\textbf{W}_{d,b}\textbf{W}_{d,b}^H\textbf{W}_a^H\textbf{E}_b^H \textbf{G}_b(\textbf{W}_a, \textbf{W}_{d,b}, \textbf{H}_b +\textbf{E}_b)^{-1} \Big) (d\textbf{E}_b)\Big]
	 \\& \quad+ {\rm Tr}\Big[ \frac{1}{B}\Big(\textbf{W}_a^* \textbf{W}_{d,b}^* \textbf{W}_{d,b}^T \textbf{W}_a^T \tilde{\textbf{H}}_b^T \textbf{G}_b(\textbf{W}_a, \textbf{W}_{d,b}, \textbf{H}_b +\textbf{E}_b)^{-T} \\
	 & \quad+
	 \textbf{W}_a^* \textbf{W}_{d,b}^*
	\textbf{W}_{d,b}^T \textbf{W}_a^T \textbf{E}_b^T \textbf{G}_b(\textbf{W}_a, \textbf{W}_{d,b}, \textbf{H}_b +\textbf{E}_b)^{-T} \Big) (d\textbf{E}_b^*)\Big]. 
	\end{align*}

Using \cite[Table 3.2]{hjorungnes2011complex}, this yields \eqref{eqn:gr_Eb_k}.

\bibliographystyle{IEEEtran}
\bibliography{IEEEabrv,refs}

\end{document}